\documentclass[12pt]{article}
\usepackage{amssymb,amsmath,graphicx}
\usepackage{bm}
\usepackage{comment}
\usepackage{dcolumn}
\usepackage[dvipdfmx]{hyperref}

\newcommand{\simgt}{\hbox{ \raise3pt\hbox to 0pt{$>$}\raise-3pt\hbox{$\sim$} }}
\newcommand{\simlt}{\hbox{ \raise3pt\hbox to 0pt{$<$}\raise-3pt\hbox{$\sim$} }}

\newcommand{\LQ}{\Lambda_{\rm QCD}}

\newcommand{\msbar}{$\overline{\rm MS}$}

\newcommand \bra[1]{\left< {#1} \,\right\vert}
\newcommand \ket[1]{\left\vert\, {#1} \, \right>}

\newcommand{\bea}{\begin{eqnarray}}
\newcommand{\eea}{\end{eqnarray}}

\newcommand{\be}{\begin{equation}}
\newcommand{\ee}{\end{equation}}
\newcommand{\fn}{\footnote}
\newcommand{\LMS}{\Lambda_{\overline{\rm MS}}}

\newcommand{\lt}{\left}
\newcommand{\rt}{\right}

\newcommand{\non}{\nonumber \\}

\newcommand{\MSbar}{\overline{\rm MS}}

\newcommand{\Vmom}{\widetilde{V}_{\rm QCD}(q)}
\newcommand{\Vmos}{\widetilde{V}_{S}(q)}
\newcommand{\Vpos}{{V}_{\rm QCD}(r)}

\begin{document}

\begin{titlepage}

    \begin{flushright}
      \normalsize TU--1094 \\
       KYUSHU--HET--205 \\
      \today
    \end{flushright}

\vskip2.5cm
\begin{center}
\Large
On renormalons of static QCD potential at $u=1/2$ and $3/2$
\end{center}

\vspace*{0.8cm}
\begin{center}
{\sc Yukinari Sumino$^a$} and
{\sc Hiromasa~Takaura$^b$}\\[5mm]
  {\small\it $^a$Department of Physics, Tohoku University,}\\[0.1cm]
  {\small\it Sendai, 980-8578 Japan}\\
  {\small\it $^b$Department of Physics, Kyushu University,}\\[0.1cm]
  {\small\it  Fukuoka, 819-0395 Japan}
\end{center}

\vspace*{2.8cm}
\begin{abstract}
\small
We investigate the $u=1/2$ 
[${\cal O}(\LQ)$]
and $u=3/2$ [${\cal O}(\LQ^3)$]
renormalons in the static QCD potential in position
space and momentum space
using the  OPE of the potential-NRQCD effective field theory.
This is an old problem and
we provide a formal formulation to analyze it.
In particular we present detailed examinations of the $u=3/2$ renormalons.
We clarify how the $u=3/2$ renormalon is suppressed in
the momentum-space potential in relation with the Wilson coefficient
$V_A(r)$.
We also point out that it is not straightforward to subtract the IR renormalon
and IR divergences simultaneously in the multipole expansion.
Numerical analyses are given,
which clarify the current status of our knowledge on
the perturbative series.
The analysis gives a positive reasoning to the method
for subtracting renormalons used in recent
$\alpha_s(M_Z)$ determination from the QCD potential.
\vspace*{0.8cm}
\noindent

\end{abstract}


\vfil
\end{titlepage}

\newpage

\section{Introduction}

For a long time the static QCD potential $\Vpos$ has been studied
extensively, in order to understand the nature of the strong interaction
between a heavy quark and antiquark pair.
In the past decades computation of $\Vpos$ in perturbative QCD
has been advanced significantly 
\cite{Fischler:1977yf,Appelquist:es,Peter:1996ig,Schroder:1998vy,Melles:2000dq,Hoang:2000fm,
Recksiegel:2001xq,Brambilla:1999qa,Kniehl:1999ud,Brambilla:2006wp,Pineda:2000gza,Brambilla:2009bi,
Smirnov:2008pn,Anzai:2009tm,Smirnov:2009fh,Lee:2016cgz}.
In association with many theoretical developments,
$\Vpos$ has become an indispensable theoretical tool to
describe not only the properties of the heavy quarkonium states
but also for precision determinations of
the fundamental parameters of QCD such as
the heavy quark masses $m_c$, $m_b$, $m_t$ 
\cite{Ayala:2014yxa,Hoang:2012us,Ayala:2012yw,Penin:2014zaa,Beneke:2014pta,
Kiyo:2015ufa,Peset:2018ria,Mateu:2017hlz,Kiyo:2015ooa,Kawabata:2016aya}
and the strong coupling constant $\alpha_s$ \cite{Bazavov:2014soa,Takaura:2018lpw}. 

Before around 1998, the prediction of $\Vpos$ in perturbative QCD 
was not successful and was plagued by the so-called
renormalon problem.
As it turned out, convergence of the perturbative series was fairly poor,
such that a meaningful prediction could not be obtained
in the distance regions relevant to the charmonium and bottomonium
states.
This is caused by the growth of $\alpha_s$
in the infrared (IR) region and is characterized by the singularities in the 
Borel transform of the perturbative series
(the singularities are called renormalons) \cite{Beneke:1998ui}.
Then it was discovered that the leading renormalon of order $\LQ$ in
$\Vpos$
is canceled
against that of the quark pole mass $m_{\rm pole}$ in the combination of the
total energy of the static quark pair $2m_{\rm pole}+\Vpos$,
which led to a dramatic improvement of convergence of
the perturbative series \cite{Pineda:id,Hoang:1998nz,Beneke:1998rk}.

Up to now, although there exists no rigorous proof
on existence of renormalons in QCD observables,
there exist standard arguments based on the
operator product expansion (OPE) and renormalization group (RG)
equations which show that their existence is consistent 
and plausible theoretically \cite{Beneke:1998ui}.
This is reinforced by a number of evidences in actual computation of
perturbative
series of QCD observables,
thanks to recent
technological developments in multiloop calculations.
There also exist examinations of the nature of renormalons
using many approximate estimates of higher-order
terms of perturbative series at various levels of rigor. See, for instance, Ref.~\cite{Bauer:2011ws}.

In many analyses of renormalons in the static QCD potential,
analyses of perturbative computation in
momentum space play important roles \cite{Beneke:1998ui}.
For $\Vpos$, it is often
assumed that there are no renormalons in its Fourier
transform $\Vmom$
(the potential in momentum space)
or that renormalons in $\Vmom$ are negligible at the current level of accuracy.
In fact, in Refs.~\cite{Hoang:1998nz} and \cite{Beneke:1998rk}, absence of 
the order $\LQ/q^3$ renormalon in $\Vmom$ (corresponding to the $u=1/2$ renormalon) 
is shown at the one-loop and two-loop levels, respectively.\footnote{
In Ref.~\cite{Beneke:1998rk}, IR divergences which arise from three loops and
beyond are neglected without a proper reasoning, and it is not clear
whether its claim is valid beyond two loop order.
}
Also, the $u=3/2$ renormalon cancellation within the multipole expansion 
was shown \cite{Brambilla:1999xf} based on the assumption of absence of 
the corresponding renormalon in $\Vmom$.
Nevertheless, it can be the case that renormalons arise from
a deep level of loop integrals in the computation of $\Vmom$
and that they are simply not detected in the currently known
several terms of the perturbative series.

A direct motivation of our study comes from necessity for a
justification for the assumption used in a recent
determination of $\alpha_s$ from $\Vpos$ \cite{Takaura:2018lpw}.
There, the first two renormalons of order 
$\LQ$ and $\LQ^3 r^2$ are subtracted from the leading-order Wilson coefficient 
of $\Vpos$,
in order to extend validity range of the OPE of $\Vpos$ to
larger $r$,
and it is assumed that 
the corresponding renormalons are negligible in the
Fourier transform of the Wilson coefficient [$\simeq \Vmom$].
This problem is also linked with how we renormalize
the IR divergences in the potential which arise from three loops and
beyond 
\cite{Appelquist:es,Brambilla:1999qa,Kniehl:1999ud,Anzai:2009tm,Smirnov:2009fh,
Brambilla:2006wp}.

In this paper we analyze the
order $\LQ$ and $\LQ^3 r^2$ renormalons in $\Vpos$,
on the basis of the standard argument by the OPE and
RG equations. 
The discussion is to a large extent based on general features of QCD,
independent of
ad hoc approximations such as the large-$\beta_0$ approximation.
We refine our understanding by looking into the detailed
structure of the OPE within the potential-NRQCD (pNRQCD)
effective field theory (EFT) \cite{Brambilla:2004jw}.
In particular we elucidate the accurate structure of the
$u=3/2$ renormalon.
Subsequently we discuss the size of the renormalon uncertainties for $u=1/2$ and $3/2$ 
in $\Vmom$ with a method which does not rely on diagrammatic analysis, 
providing a different perspective from, e.g., 
Refs.~\cite{Beneke:1998rk,Brambilla:1999xf}.
We also believe that an argument such as the one we provide in Sec.~3
is necessary to clarify treatment of the IR part of $\Vmom$.
In the latter part of this paper, 
we test our understanding by performing
numerical analyses of the normalization constants of the renormalons 
in the perturbative series of $\Vpos$ and $\Vmom$.
We treat two kinds of perturbative series:
one is the fixed-order perturbative series currently known 
and the other includes higher-order terms estimated by RG. 
We estimate the normalization constants of the renormalons from these
two perturbative series by using Lee's method \cite{Lee:1996yk} and 
also by an analytic formula which we derive
 (for the latter series).
The former includes updates of the analyses by Pineda \cite{Pineda:2001zq}.
See also Ref.~\cite{Ayala:2014yxa} for a more recent result.

The paper is organized as follows.
In Sec.~2 we briefly review the standard argument
on renormalons for a general QCD observable.
In Sec.~3 we scrutinize the structure of renormalons
in the static QCD potential.
Secs.~4--8 present numerical analyses on the normalization constants
of the renormalons.
In Sec.~4 we study the ${\cal O}(\LQ)$ renormalon of $\Vpos$
from fixed-order perturbative series,
followed by a study of its cancellation with that of the pole mass
in Sec.~5.
We compare these results with that obtained by an integral
formula in Sec.~6.
We study the ${\cal O}(\LQ^3r^2)$ renormalon of $\Vpos$
in Sec.~7.
Finally we test the corresponding renormalons
in $\Vmom$ in Sec.~8.
Conclusions are given in Sec.~9.
In App.~A we explain theoretical aspects of IR cancellation
at $O(r^0)$ of the multipole expansion.
In App.~B we present details of the derivation of a formula for
the normalization of renormalons in $\Vpos$.


\section{Structure of renormalons}
\label{sec:Theory}

Let us first review briefly the structure of renormalons 
in QCD observables \cite{Beneke:1998ui}.

Consider a general RG-invariant dimensionless observable $X(Q)$ with a
typical energy scale $Q$.
Its perturbative expansion is given by
\bea
X^{\rm PT}(Q) = \sum_{n=0}^{\infty} d_n(Q/\mu) \alpha_s(\mu)^{n+1} \,.
\label{pX}
\eea
$\mu$ denotes the renormalization scale in the {\msbar} scheme.
It satisfies the RG equation
\bea
\mu^2\frac{d}{d\mu^2}X^{\rm PT}(Q)=
\left[ \mu^2 \frac{\partial}{\partial \mu^2}+
\beta(\alpha_s)\frac{\partial}{\partial \alpha_s} \right] 
X^{\rm PT}(Q)=0
\eea
with the beta function given by
\bea
\mu^2 \frac{d \alpha_s}{d \mu^2}=\beta(\alpha_s)=-\sum_{i=0}^\infty b_i \, \alpha_s^{i+2}
\label{MSbarRGeq-alfs}
\,.
\eea
The first two coefficients of the beta function are given explicitly by
\be
b_0=\frac{1}{4 \pi} \lt(11-\frac{2}{3} n_f \rt),
~~~~~
b_1=\frac{1}{(4 \pi)^2} \lt(102-\frac{38}{3} n_f \rt) \, .
\ee

It is conjectured that for many observables 
the coefficients of the perturbative series grow factorially,
$d_n \sim n!$, for large $n$.
To quantify uncertainties induced by
this property, the Borel transform of $X^{\rm PT}$, defined by
\be
B_X(t)=\sum_{n=0}^{\infty} \frac{d_n}{n!} t^n \, ,
\ee
is studied.
Renormalons of $X^{\rm PT}$ refer to the singularities of $B_X(t)$
located on the real axis in the complex $t$-plane.
We assume the form of the Borel transform in the vicinity of each renormalon singularity at $t=u/b_0$
as
\begin{align}
&B_X(t)=
\left( \frac{\mu^2}{Q^2} \right)^u 
\frac{N_{u}}{\displaystyle \bigl(1- {b_0 t}/{u}\bigr)^{1+\nu_u} } \, 
\sum_{k=0}^\infty c_k(\mu/Q) \left(1- \frac{b_0 t}{u}\right)^k
+\text{(regular part)}
\,,
\label{AnalyFormBX}
\\
&c_0=1
\,,
\end{align}
with parameters $N_u$, $\nu_u$ and $c_k$'s.
This form is consistent with the RG equation.
Formally we can reconstruct $X^{\rm PT}$ from
its Borel transform $B_X(t)$ by the inverse Borel transform given by 
the integral
\be
X^{\rm PT}``="\int_0^{\infty} dt \,B_X(t) e^{-t/\alpha_s(\mu)} \, . \label{invBorelTr}
\ee
However, if there are singularities (renormalons) on the positive
real axis, the integral is ill defined.
We can regularize the integral by deforming the integral contour to
the upper or lower half plane:
\be
X_{\pm}^{\rm PT}=\int_0^{\infty \times \exp(\pm i \epsilon)} dt \, B_X(t) e^{-t/\alpha_s(\mu)}\,.
\ee
We can define the ambiguity induced by the renormalon from the discontinuity of the corresponding singularity,
and the singularity at $t=u/b_0$ with $u>0$ of eq.~\eqref{AnalyFormBX} gives 
\begin{align}
&{\rm Im}\,X^{\rm PT}_{\pm}[u]=
\pm \frac{\pi}{b_0} \frac{N_u}{\Gamma(1+\nu_u)} u^{1+\nu_u} \lt(\frac{\LMS^2}{Q^2} \rt)^u
(b_0 \alpha_s(Q))^{-\nu_u+u b_1/b_0^2} \sum_{k=0}^{\infty} \tilde{c}_k \alpha_s(Q)^k \, ,\non
& ~~\tilde{c}_0=1 \label{ambX} \,,
\end{align}
where we have used
\bea
\LMS^2=\mu^2 \, \exp\left[-\left\{
\frac{1}{ b_0 \alpha_s}+\frac{b_1}{b_0^2}\log(b_0\alpha_s)
+\int_0^{\alpha_s}\! dx\,
\left( \frac{1}{\beta(x)}+\frac{1}{b_0x^2}-\frac{b_1}{b_0^2x}\right)
\right\}\right]
\,.
\eea

The parameters $u$, $\nu_u$, $c_k$ and $\tilde{c}_k$
in eq.~\eqref{AnalyFormBX} or eq.~\eqref{ambX}  can usually be determined from the OPE.
In the context of the OPE in $1/Q$, $X^{\rm PT}$
is identified with the Wilson coefficient of the leading 
identity operator.
Let us denote by $O_u$ the lowest dimension (dimension $2u$) renormalized
operator responsible for cancellation of the leading renormalon 
in $X^{\rm PT}$.
For simplicity we discuss the case where only one operator
is involved.
The OPE reads
\bea
&&
X(Q)=C_1^X(Q)+C^X_{O_u}(Q/\mu, \alpha_s(\mu))\frac{\bra{0}O_u(\mu)\ket{0}}{Q^{2u}}
+\cdots  \label{OPE}
\,,
\\&&
C^X_1(Q)=X^{\rm PT}(Q) 
\,, ~~~C^X_{O_u}(Q/\mu,\alpha_s(\mu))=\sum_{n=0}^{\infty} f_n(Q/\mu) \alpha_s(\mu)^n
\,.
\eea
We assume that the leading ambiguity induced by the renormalon
of $C_1^X$ as given in eq.~\eqref{ambX} is canceled by the second term of the OPE.
Then, the $Q$-dependence of the renormalon uncertainty of $C_1^X$ 
should coincide with that of the second term in the OPE, which can be detected as follows.
Suppose that the Wilson coefficient satisfies the RG equation,
\be
\lt[\mu^2 \frac{d}{d \mu^2} -\gamma(\alpha_s) \rt] C_{O_u}^X=0~~~;~~~
\gamma(\alpha_s)=\sum_{i=0}^\infty \gamma_i \, \alpha_s^{i+1}
\,.
\ee
This RG equation specifies the $Q$-dependence of $C^X_{O_u}(Q/\mu,\alpha_s(\mu))$ as
\begin{align}
C_{O_u}^X(Q/\mu,\alpha_s(\mu))
&=\exp \lt[-\int_{\alpha_s(\mu)}^{\alpha_s(Q)}dx \, \frac{\gamma(x)}{\beta(x)} \rt] C_{O_u}^X(1,\alpha_s(Q)) \non
&=const. \times 
[\alpha_s(Q)]^{\gamma_0/b_0} [1+\mathcal{O}(\alpha_s(Q))] C_{O_u}^X(1,\alpha_s(Q)) \, , \label{COu}
\end{align}
where $const.$ denotes a $Q$-independent (but $\mu$-dependent) constant.
Now the $Q$-dependence of the second term of the OPE \eqref{OPE} is made explicit.

Requiring the same $Q$-dependence for the renormalon uncertainty of $X^{\rm PT}$,
using eq.~\eqref{OPE} with eq.~\eqref{COu}, we obtain for eq.~\eqref{AnalyFormBX} or \eqref{ambX} 
\be
\nu_{u}=\frac{b_1}{b_0^2}u -\frac{\gamma_0}{b_0}\, . \label{nuu}
\ee
The factor $\sum_{k=0}^{\infty} \tilde{c}_k \alpha_s(Q)^k$ in eq.~\eqref{ambX} 
should be proportional to
$[1+\mathcal{O}(\alpha_s(Q))] C_{O_u}^X(1,\alpha_s(Q))$ in eq.~\eqref{COu}.  
Therefore, $c_k$'s and $\tilde{c}_k$'s 
can be determined one by one
from smaller $k$ in terms of $b_n$'s, $\gamma_n$'s and $f_n$'s 
from smaller $n$.
The overall normalization $N_u$ cannot be determined from this argument.
We note that, in the case that $C^X_{O_u}O_u$ is
independent of $Q$, $\gamma_0=0$ and $\tilde{c}_k=0$ for $k\ge 1$.


\section{Renormalons in the static QCD potential}

In this  section we investigate theoretical aspects of
renormalons of the static QCD potential,
focusing on the $u=1/2$ and $3/2$ renormalons, on the basis
of the above general understanding.
Part of the argument given in this section has already been
discussed in \cite{Brambilla:1999qa}.
(See also \cite{Sumino:2014qpa}.)
We refine the discussion and present new observations.
In particular, main part of the discussion on the $u=3/2$ renormalon is new.

\subsection{Basics of static QCD potential}
\label{sec:3.1}
The static QCD potential
is defined from an expectation value
of the Wilson loop as
\be
V_{\rm QCD}(r)
=
- \! \lim_{T \to \infty} \frac{1}{iT} \,
\log \frac{\bra{0} {\rm{Tr\, P}}
\exp\left[ i g \oint_{C} dx^\mu A_\mu(x) \right]
\ket{0}}
{\bra{0} {\rm Tr} \, {\bf 1} \ket{0}}
\,,
\label{defalphaV}
\ee
where 
${C}$ is a rectangular loop of spatial extent $r$ and
time extent $T$.
P stands for the path-ordered product along the contour $C$.
It is conjectured that renormalon singularities are located
in the Borel transform of the perturbative
series of $\Vpos$ at
$t=\frac{1}{2 b_0},\, \frac{3}{2 b_0}$, etc.\
(i.e., $u=\frac{1}{2}, \, \frac{3}{2}$, etc.).

In calculation of the static QCD potential, we have two different scales. One is the soft scale $1/r$, which is the 
inverse of the distance between a static $Q \bar{Q}$ pair.
The other is the ultrasoft (US) scale, which is set by the energy difference of the
color singlet and octet states of the static $Q\bar{Q}$ pair,
\bea
\Delta V(r)=V_O(r)-V_S(r)
\,,
\eea
where
\bea
&&
V_S(r)=-C_F\frac{\alpha_s}{r} + {\cal O}(\alpha_s^2)
,
\label{V_S}
\\ &&
V_O(r)=\Bigl(\frac{C_A}{2}-C_F\Bigr)
\frac{\alpha_s}{r} + {\cal O}(\alpha_s^2)
,
\label{V_O}
\\ &&
\Delta V(r)=\frac{C_A}{2}
\frac{\alpha_s}{r} + {\cal O}(\alpha_s^2)
\,.
\label{Pert-DeltaV}
\eea
The pNRQCD EFT describes dynamics in which the $Q\bar{Q}$ system 
emits or absorbs US gluons whose energies are comparable to or smaller 
than the energy differences of different $Q\bar{Q}$ states \cite{Brambilla:1999xf}.
Accordingly, the factorization scale $\mu_f$ (= cut off scale of pNRQCD) is chosen to satisfy
$\Delta V \ll \mu_f \ll 1/r$.

The OPE of the static QCD potential $V_{\rm QCD}(r)$ in $r$ can be performed 
within the pNRQCD EFT in the static limit based on the scale hierarchy
$1/r \gg \Delta V$:\footnote{
This is equivalent to 
$\frac{1}{2}C_A\alpha_s(1/r)\ll 1$, which holds at
sufficiently small $r$.}
\be
V_{\rm QCD}(r)=V_S(r)+\delta E_{\rm US}(r) +\dots \, . \label{VQCDinpNRQCD}
\ee
The leading term $V_S(r)$ denotes the singlet potential,
which is the Wilson coefficient of the bilinear singlet field operator
$S^\dagger S$ in the context of pNRQCD.
In eq.~\eqref{VQCDinpNRQCD}, $V_S(r)$ is multiplied by $\bra{S}S^\dagger S\ket{S}=\bra{0}{\bf 1}\ket{0}=1$.
The second term is the ${\cal O}(r^2)$ term in the
multipole expansion, given by
\bea
\delta E_{\rm US}(r)=
-i  \, \frac{V_A(r)^2}{6}\!\int_0^\infty \!\!\! dt
\, e^{-i \,t \, \Delta V(r)}
\langle\,g\,\vec{r}\!\cdot\! \vec{E}^a(t,\vec{0})\,
\varphi_{\rm adj}(t,0)^{ab}\,
g\,\vec{r}\!\cdot\! \vec{E}^b(0,\vec{0})\,\rangle 
\,,
\label{MatchingRel}
\eea
where
$\vec{E}=-\partial_t \vec{A}-\vec{\partial}A_0-ig[A_0,\vec{A}]$
represents the color electric field;
the color string for the adjoint representation is given by
$
\varphi_{\rm adj}(t,t')=
{\rm T}\, \exp\Bigl[ig\int_{t'}^t d\tau\,
A_0^c(\tau,\vec{X})\,T^c_\text{adj}
\Bigr]
$.
$\delta E_{\rm US}$ is generated by insertions of
the operators $g\, O^{a\dagger}\vec{r}\cdot \vec{E}^a S$
and $g\, S^\dagger \vec{r}\cdot \vec{E}^a O^a$,
and $V_A(r)$ denotes the Wilson coefficient of these operators.
Note that eqs.~\eqref{VQCDinpNRQCD} and \eqref{MatchingRel}
are exact to all orders in $\alpha_s$.

$V_S(r)$ coincides with
the naive expansion of $V_{\rm QCD}(r)$ in $\alpha_s$:
\bea
V_S(r) = V_{\rm QCD}(r)\Bigr|_\text{exp.\ in $\alpha_s$}
.
\label{V_SfromV_QCD}
\eea
To see this, we adopt the energy integral representation of $\delta E_{\rm US}(r)$,
\be
\delta E_{\rm US}(r)=
- \,  \, \frac{V_A(r)^2}{6}
r_i r_j \, \int_{-\infty}^\infty \! \frac{dk}{2\pi}
\, \frac{1}{k+\Delta V(r)} \, 
\langle\,{gE}^a_i\,
\varphi_{\rm adj}^{ab}\,
{gE}^b_j\,\rangle(k) 
\,,
\label{deltaEUS-energyintegral}
\ee
which can be obtained with Fourier transform of the correlation function,
\be
\langle\,{gE}^a_i\,
\varphi_{\rm adj}^{ab}\,
{gE}^b_j\,\rangle(k) \equiv
\int_{-\infty}^\infty \! dt \, e^{ikt}\,
\langle\,{gE}^a_i(t,\vec{0})\,
\varphi_{\rm adj}(t,0)^{ab}\,
{gE}^b_j(0,\vec{0})\,\rangle
\,.
\ee
Then, if we naively expand $\delta E_{\rm US}$ in $\alpha_s$
before loop integrations, the US scale $\Delta V=\mathcal{O}(\alpha_s)$ disappears from the
propagator denominator in eq.~\eqref{deltaEUS-energyintegral}, 
and the integrals become scaleless and vanish.
(The same applies to beyond ${\cal O}(r^2)$ terms.)
We can rephrase this in the computation of $\Vpos$ 
in expansion in $\alpha_s$, by applying
expansion-by-regions technique to loop integrals \cite{Beneke:1997zp}. 
We can separate contributions
from the UV scale $1/r$ and the US scale ($\ll 1/r$), where
the latter contributions vanish to all orders in $\alpha_s$ since
they are given by scaleless integrals.


We investigate theoretical aspects of renormalons at $u=1/2$ (Sec.~\ref{sec:3.2})
and $u=3/2$ (Sec.~\ref{sec:3.3}) based on the above general understanding, and in particular determine some of the parameters in 
eq.~\eqref{AnalyFormBX} assuming this expansion form
around the singularities.
Before this, let us comment on the IR divergences present in the perturbative result of $V_S(r)$.
The naive perturbative expansion of $\Vpos$ includes IR divergences at and beyond order $\alpha_s^4$
\cite{Appelquist:es,Brambilla:1999qa,Kniehl:1999ud,Anzai:2009tm,Smirnov:2009fh,
Brambilla:2006wp}, hence so does $V_S(r)$.
The IR divergences of $V_S(r)$ have their counterparts in the OPE at
order $r^2$ or beyond in the multipole expansion. 
Indeed, $\delta E_{\rm US}$ contains UV divergences if we
compute it in double expansion in $\alpha_s$
and $\log\alpha_s$ consistently with the philosophy
of pNRQCD, that is, keeping $\Delta V$[$\simgt k$: gluon momentum in Eq.~\eqref{deltaEUS-energyintegral}]
in the propagator denominator.
At ${\cal O}(r^2)$, the UV divergences of $\delta E_{\rm US}$
and IR divergences of $V_S(r)$ cancel in $\Vpos$, reflecting the $\mu_f$-independence of $\Vpos$.
In the subsequent argument, we implicitly assume a certain regularization 
prescription for making these divergences finite
to discuss renormalons in the perturbative series whose each 
expansion coefficient is finite.
We will propose explicit regularization (renormalization) schemes 
and also discuss their relevance to the renormalon structure (Sec.~\ref{sec:3.5})
after the renormalon structure is clarified.

\subsection{$u=1/2$ renormalon}
\label{sec:3.2}
\begin{figure}[t]
\begin{minipage}{0.5\hsize}
\begin{center}
\includegraphics[width=4cm]{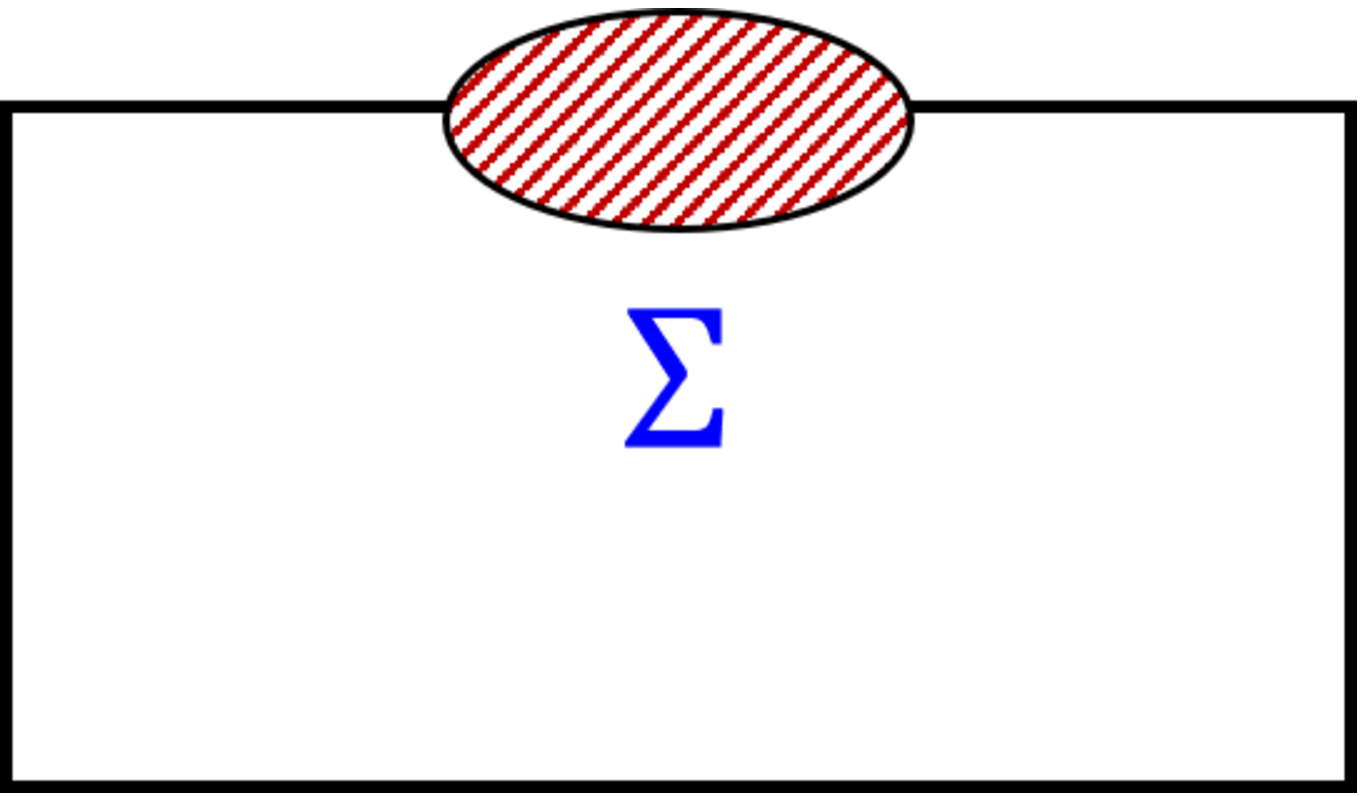}
\end{center}
\end{minipage}
\begin{minipage}{0.5\hsize}
\begin{center}
\includegraphics[width=4cm]{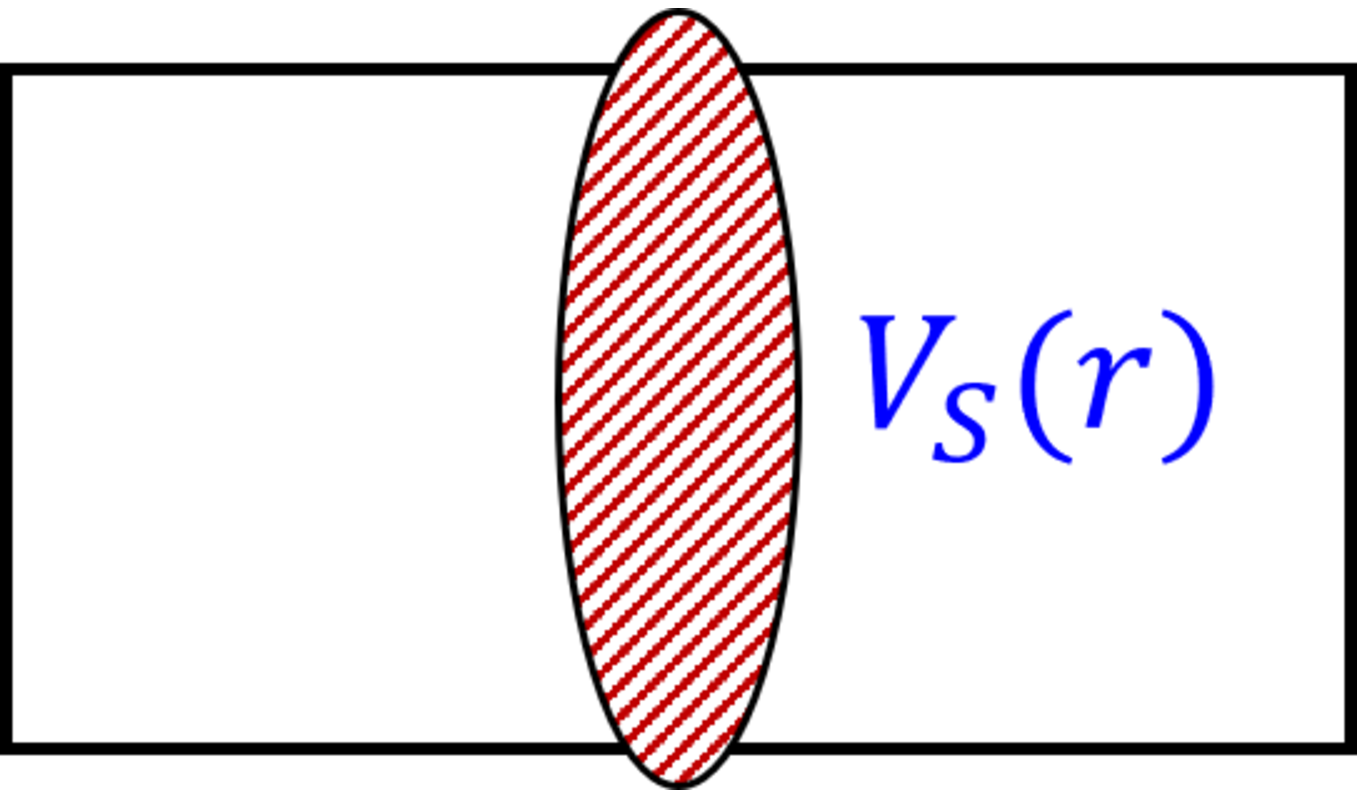}
\end{center}
\end{minipage}
\caption{\small
Schematic representations of diagrams contributing to $\Sigma$ and $V_S(r)$.
\label{Fig:Sigma+Vs}
}
\end{figure}

Let us clarify the current understanding on the $u=1/2$ renormalon. 
The leading IR renormalon of $V_S(r)$ is located at $u=1/2$, and
the induced ambiguity is known to be independent of $r$ and proportional to $\LMS$.
In fact, the $r$-independent constant part of
$\Vpos$ in eq.~\eqref{defalphaV} is not well defined.
This is inherent in the self-energy type contributions $\Sigma$ to each
static color charge.
These contributions vanish in perturbative computation in dimensional
regularization, since they are given by scaleless integrals.
Hence, they are not included in the computation of $V_S(r)$,
which consists of
the potential-energy type contributions (represented by diagrams with
cross talks between the two static charges).
See Fig.~\ref{Fig:Sigma+Vs}.
The IR contributions to the self-energies $2\Sigma$ cancel 
against the IR contributions to $V_S(r)$.
This is represented in pNRQCD by the absence of ${\cal O}(r^0)$
interactions of the singlet field and US gluon field
and is a consequence of the fact that in the IR limit
gauge field couples to the total charge ($=0$ for $\ket{S}$);
further explanation is given in App.~A.
On the other hand, $\Sigma$ is UV divergent, and
in dimensional regularization simply $\Sigma$ is set equal to zero.
Thus, more precisely, $V_S(r)$ should be written as
$2\Sigma + V_S(r)$ in eq.~\eqref{VQCDinpNRQCD}, but $\Sigma$
is omitted in accordance with the usual convention.

A standard way to confirm cancellation of the $r$-independent
IR contributions to $V_S(r)$ with the self-energy type contributions
is to show cancellation of the $u=1/2$ renormalons in the combination
$2m_{\rm pole}+V_S(r)$ \cite{Pineda:id,Hoang:1998nz,Beneke:1998rk}.
By construction of pNRQCD as a low energy EFT, 
IR contributions to $\Sigma$ and $m_{\rm pole}$ are common.
Both $\Sigma$ and $m_{\rm pole}$ are RG invariant, hence
ambiguities induced by the leading renormalons both correspond
to $u=1/2$ and proportional to $\LMS$.
This reasoning determines the parameters
for the dimensionless 
potential $r V_S(r)$  in eq.~\eqref{ambX} and 
consequently in eq.~\eqref{AnalyFormBX} as
\be
\nu_{1/2}=\frac{b_1}{2 b_0^2}  \label{nuonehalf}
\ee 
and
\be
\tilde{c}_k=0 \quad \text{for $k \geq 1$} \, .
\ee

\subsection{$u=3/2$ renormalon}
\label{sec:3.3}
To clarify the structure of the $u=3/2$ renormalon,
the $r$-dependence of the $u=3/2$ renormalon uncertainty 
should be revealed (as done in Sec.~\ref{sec:Theory}). 
To this end, we focus on $\delta E_{\rm US}(r)$ [for instance, the expression of eq.~\eqref{deltaEUS-energyintegral}],
which cancels the corresponding renormalon uncertainty of $V_S(r)$.
At this stage, we note that computation of $V_S(r)$ does not include the US scale, and thus
the $u=3/2$ renormalon uncertainty is independent of $\Delta V$. 
(Supplementary discussion on this point is given in Sec.~\ref{sec:3.5}.)
This reasoning and the expression, for instance, of eq.~\eqref{deltaEUS-energyintegral}
tell us that the $r$-dependence of the $u=3/2$ renormalon uncertainty is given solely by $\sim r^2 V_A^2(r)$.
Now we investigate the $r$-dependence of $V_A(r)$.
Since $V_A(r)$ can be renormalized multiplicatively,
the RG equation of the form $[\mu^2 d/(d \mu^2)-\gamma(\alpha_s)] V_A(r)=0$ follows,\fn{
Here we are concerned with the logarithms associated with the UV
divergences of $V_A(r)$ in the full theory (or with
respect to the soft scale).
This RG equation with respect to $\mu$
 is different from the RG equations with respect to $\mu_f$
considered in Refs.~\cite{Pineda:2000gza,Brambilla:2009bi,Pineda:2011aw},
which are associated with the IR divergences with respect to
the soft scale. See discussion in Sec.~\ref{sec:3.5}.
} 
where $\gamma(\alpha_s)=\gamma_0 \alpha_s+\gamma_1 \alpha_s^2+\dots$.
From this RG equation, the fixed order result of $V_A(r)$ takes the form
\be
V_A(r;\mu)=e_0+(e_1+e_0 \gamma_0 \log(\mu^2 r^2)) \alpha_s(\mu)+\mathcal{O}(\alpha_s^2) \, .
\ee
From the explicit NLO result $V_A(r)=1+\mathcal{O}(\alpha_s^2)$ \cite{Pineda:2000gza}, we see that $\gamma_0=0$.
Thus, we determine the parameter $\nu_{3/2}$ for the dimensionless 
 potential in the Borel transform
 in eq. (6) as [cf. eq. (16)]
\be
\nu_{3/2}=\frac{3 b_1}{2 b_0^2}  \, . \label{nuthreehalf}
\ee
We also clarify that the $u=3/2$ renormalon uncertainty is given by eq.~\eqref{ambX} with $(b_0 \alpha_s(1/r))^{-\nu_u+u b_1/b_0^2}=1$.\fn{
We implicitly assume that the $u=3/2$ renormalon uncertainty is RG invariant as we assume eq.~\eqref{ambX}, 
although RG invariance of $V_S(r)$ may be violated by the IR divergences
(or IR logarithms).
This assumption is justified when we adopt explicit schemes 
to remove the IR divergences from $V_S(r)$
such that the redefined $V_S$ is RG invariant;
see Sec.~\ref{sec:3.5}. \label{fn:RGinv}
}
The parameters $\tilde{c}_k$ can be parametrized by $e_i$'s, $\gamma_i$'s, and $b_i$'s.
With the NLO result of $V_A$, $\tilde{c}_1$ is explicitly obtained as
\be
\tilde{c}_1=-\gamma_1/b_0 \, . \label{c1tildethreehalf}
\ee
Thus, a correction factor to the $r$-dependence of the renormalon uncertainty $r^2$ is 
given by $1+\mathcal{O}(\alpha_s(1/r))$ with the above $\tilde{c}_1$ term.
(The explicit result of $\gamma_1$ is not known currently.)

\subsection{Renormalon in momentum-space potential}
\label{sec:3.4}
We now discuss the renormalon uncertainty in the momentum-space potential.
For convenience we define the dimensionless potential 
as $v(r)=r V_{S}(r)$.
Suppose that
we have the ambiguity  in the position-space potential due to 
the renormalon at $t=u/b_0$ as
\be
{\rm Im}\, {v}_{\pm}[u]= \pm N_{u} A(u) \, (r^2 \LMS^2)^u  \, , \label{amb1}
\ee
with
\bea
A(u)=\frac{\pi}{b_0} \frac{1}{\Gamma(1+u b_1/b_0^2)}  u^{1+u \frac{b_1}{b_0^2}}
\,.
\eea
For $u=1/2$, eq.~\eqref{amb1} is exact, 
whereas the correction factor $1+\mathcal{O}(\alpha_s(1/r))$ arises for $u=3/2$. 
The momentum-space potential is obtained by the Fourier transform of
$V_S(r)$:
\be
\Vmos =-4 \pi C_F \frac{\widetilde{\alpha}_V(q)}{q^2} = \int d^3 \vec{r} \, e^{-i \vec{q} \cdot \vec{r}} \, V_S(r) \, .
\ee
Here and hereafter, we denote $\Vmos$ instead of $\Vmom$
to make explicit that we consider the Fourier transform of the leading Wilson
coefficient $V_S(r)$, although in naive perturbative expansion $V_S(r)$ and
$\Vpos$ coincide; see eq.~\eqref{V_SfromV_QCD}.

From the Fourier transform of $v_\pm$, we can obtain 
the renormalon uncertainty in the
$V$-scheme coupling constant in momentum space:
\be
{\rm Im}\, \widetilde{\alpha}_V(q)_\pm [u] =\mp\frac{N_u}{C_F} A(u) \lt( \frac{\LMS}{q} \rt)^{2 u} \Gamma(2u+1) \cos ({\pi u}) \, ,
\label{Vmom-renorm-absence}
\ee
where we have used analytical continuation  of the result for $(-1<)u<0$.
The above formula shows that, if we
assume eq.~\eqref{amb1}, renormalons of $\Vmos$ vanish 
at positive half-integer $u$'s, since $\cos ({\pi u})=0$ and $A(u)$ is finite.
In particular, the normalization of the renormalon at $u=1/2$ vanishes,
\be
{\rm Im}\, \widetilde{\alpha}_V(q)_\pm [u=1/2]=0 \, .
\ee

For $u=3/2$, while the normalization is not exactly zero,
it is suppressed by $\alpha_s(q)^2$. To see this, one should
note first eq.~\eqref{nuthreehalf} in eq.~\eqref{ambX},
secondly that $\tilde{c}_k$'s are independent of $1/r$ and also
$\alpha_s(1/r)=\alpha_s(q)+b_0 \log(r^2 q^2) \alpha_s(q)^2+\cdots$.
Explicitly, the leading behavior of the $u=3/2$ renormalon uncertainty of $\alpha_V(q)$ is given by\fn{
The uncertainty \eqref{Vmom-renorm-suppress} is obtained in a parallel form to eq.~\eqref{ambX}
in the sense that the part given by the series expansion in $\alpha_s$ is specified with $b_i$'s and $\gamma_i$'s.
Thus, the result sounds plausible.
}
\be
{\rm Im}\, \widetilde{\alpha}_V(q)_\pm [u=3/2] 
= \mp\frac{N_{3/2}}{C_F} A(u=3/2) \lt( \frac{\LMS}{q} \rt)^{3} \alpha_s(q)^2 
\, 6 \pi  b_0  \tilde{c}_1 [1+\mathcal{O}(\alpha_s(q))] \, ,
\label{Vmom-renorm-suppress}
\ee
where $N_{3/2}$ and $\tilde{c}_1$ represent the parameters of the position-space potential.

Thus, eqs.~\eqref{Vmom-renorm-absence}--\eqref{Vmom-renorm-suppress} provide a formal framework to
analyze renormalons in the momentum-space potential, without recourse
to diagrammatic analyses (or resummation of certain diagrams)
used in the previous analyses \cite{Hoang:1998nz,Beneke:1998rk,Brambilla:1999qa}.

\subsection{Renormalization scheme}
\label{sec:3.5}
So far, we did not specify how to renormalize $V_S(r)$ and $\delta E_{\rm US}(r) $,
which contain the IR divergences and UV divergences, respectively,
Here, we define two schemes to remove the divergences.
\medbreak\noindent
\underline{Scheme (A)}
\medbreak

At each order of the perturbative expansion of ${V}_S(r)$ in $\alpha_s$, 
we first set $\mu=1/r$ and then
drop all the poles in $\epsilon$ originating from the IR divergences.
($\mu$ denotes the renormalization scale in
full QCD. The scale in the argument of the logarithms originating from
the IR divergences is also taken as $\mu$,  even though
it is sometimes distinguished in the literature.)
We also redefine $\delta E_{\rm US}$ such that  the sum
$V_S(r)+\delta E_{\rm US}$ is unchanged,  which is evaluated
in double expansion in $\alpha_s$ and $\log\alpha_s$.
The renormalized $V_S$ and
$\delta E_{\rm US}$ are both independent of $\mu$ by definition.\footnote{
$V_S(r)$ and $\delta E_{\rm US}(r) $ at different $\mu$
are obtained by rewriting $\alpha_s(1/r)$ by $\alpha_s(\mu)$.
}

In fact, this regularization is compatible with the property used in Sec.~\ref{sec:3.3} that
$V_S(r)$ is RG invariant (see footnote \ref{fn:RGinv}), but it may be incompatible with the one that 
the $r$-dependence of the $u=3/2$ renormalon uncertainty is given by $\sim r^2 V^2_A(r)$.
This is because the latter reasoning [and thus the results such as eq.~\eqref{nuthreehalf}]
relies on eq.~\eqref{deltaEUS-energyintegral} and additional contribution was not considered.
However, we assume that the structure of IR renormalons in $V_S(r)$ at ${\cal O}(r^2)$
is unchanged by this prescription.
This is indeed the case in the large-$\beta_0$ approximation
of $\delta E_{\rm US}$,
in which the IR divergences and IR renormalons are clearly separated;
the former is given as a convergent series in $\alpha_s$
expansion, while the latter is given as a factorially diverging series.
This is shown by computing $\delta E_{\rm US}$ in the large-$\beta_0$ approximation \cite{Sumino:2004ht}:
\bea
\delta E_{\rm US}(r) \Bigr|_\text{large-$\beta_0$}
&=&
\frac{C_F\alpha_s}{4\pi}\,8r^2\, \Delta V(r)^3
\sum_{n=0}^\infty \left({b_0\alpha_s}\right)^n
\left[ n! \, G_{n+1}+
\frac{1}{\epsilon^{n+1}}\,\frac{(-1)^n}{n+1}\, g(\epsilon)\right]
\nonumber\\
&& 
+\,{\cal O}(\epsilon,r^3)  \label{deltaEUSlargebeta0}
\eea
where
\bea
&&
G(u)\equiv \sum_{j=0}^\infty G_j u^j
=
\left[ \frac{\mu_f\,e^{5/6}}{2\, \Delta V(r)} \right]^{2u} \,
\frac{2\,\Gamma (2-u)\, \Gamma (2u-3)}{\Gamma(u-1)} \,,
\label{GofdeltaEUS}
\\
&&
g(\epsilon)=\frac{\Gamma(4-2\epsilon)}
{36\, \Gamma(1+\epsilon)\,\Gamma(2-\epsilon)^2\,\Gamma(1-\epsilon)}
\,.
\eea
The UV divergences and UV renormalons of $\delta E_{\rm US}(r)$
are canceled, respectively, by the IR divergences and IR renormalons
of $V_S(r)$.
In addition, the US logarithms at LL \cite{Pineda:2000gza} and NLL
\cite{Brambilla:2009bi,Pineda:2011aw}, associated with the
IR divergences in $V_S(r)$, are known to be given by
convergent series in $\alpha_s$ expansion, computed explicitly using
the RG equation of pNRQCD.
Thus, to the best of our knowledge, the above assumption seems to be valid.
As a result, we consider that the scheme (A) is suitable for 
studying the renormalon of $V_S(r)$,
where the renormalon structure revealed in Sec.~\ref{sec:3.3}--\ref{sec:3.4} based on the OPE 
is not expected to be modified.

We note the existence of the UV renormalons at 
$u=3/2$, $1/2$, $-1/2$, $-3/2$, $\cdots$ of $\delta E_{\rm US}$ in eq.~\eqref{GofdeltaEUS}.
It is confirmed that the leading UV 
renormalon at $u=3/2$ cancels the known ${\cal O}(r^2)$ IR renormalon
of $V_S(r)$ in the large-$\beta_0$ approximation \cite{Sumino:2004ht, Takaura:2017lwd}.
The subleading renormalon at $u=1/2$ is also expected to be canceled against $V_S(r)$\footnote{
The poles on the negative axis are not problematic
since they are Borel summable.
} 
(although it cannot be confirmed within the large-$\beta_0$ approximation)
since the IR structure of $V_S(r)$ should match the UV structure of $\delta E_{\rm US}$.
The residues of the subleading renormalons at smaller $u$ are
proportional to higher powers of $\Delta V(r)/\mu_f$.
This leads to less powers\footnote{
Since $\Delta V(r) \sim (r |\log r|)^{-1}$, the form
of the renormalons are not integer powers of $r$.
\label{fn:10}
}
of $r$, which contradicts to the naive expectation that the renormalons
of $V_S(r)$
beyond $u=3/2$ are suppressed by higher powers of $r$ in accordance
with the multipole expansion.
This feature originates from the fact that if we expand
the integrand of 
eq.~\eqref{deltaEUS-energyintegral} in $\Delta V$,
higher power singular IR behaviors $\sim (\Delta V(r)/k)^n$ appear.
[Note that $\langle\,{E}^a_i\,
\varphi_{\rm adj}^{ab}\,
{E}^b_j\,\rangle(k) $
is independent of $\Delta V$.]
The IR structure of
$V_S(r)$ includes the same power behaviors, 
since the IR structure of $\delta E_{\rm US}$
is common to that of $V_S(r)$ once the integrand is expanded in $\alpha_s$.
The higher power singular IR behaviors generate the above more singular
IR renormalons as well as higher power IR divergences.\footnote{
Up to date, these more singular IR renormalons have not been
investigated seriously.
One reason would be that they are generated only at higher loops,
since they arise with higher powers of $\Delta V$.
In this connection, we note that the UV renormalons at $u<3/2$ in 
$\delta E_{\rm US}|_\text{large-$\beta_0$}$ do not have their
IR renormalon counterparts in $V_S|_\text{large-$\beta_0$}$
because the order counting is different between these quantities.
The former are suppressed by higher powers
of $\Delta V\sim{\cal O}(\alpha_s)$ compared to the latter.
We need to go beyond the large-$\beta_0$ approximation
to detect these IR renormalons in $V_S(r)$.
}

The above observation in particular means that
$V_S(r)$ has a renormalon at $u=1/2$ corresponding to the above UV renormalon of $\delta E_{\rm US}$.
The $u=1/2$ renormalon uncertainty is given by $\mathcal{O}(\LMS r^2 \Delta V(r)^2)$,
as seen from eqs.~\eqref{deltaEUSlargebeta0} and \eqref{GofdeltaEUS}.
It is similar to the form which is derived by the RG equations 
of the US scale $\mu_f$ on the assumption that  $\mathcal{O}(r^0)$
terms (up to anomalous dimensions) are contained in $V_S(r)$ \cite{Brambilla:2009bi}.
In particular, a $\mu_f$-independent term $\sim \LMS r^2 \Delta V(r)^2$, which is of the same form
as the renormalon uncertainty, can be added to 
Eq.~(25) in Ref.~\cite{Brambilla:2009bi} without spoiling the solution to the RG equations
(or we can say that this term is included in the $\mu_f$-independent but possibly $r$-dependent 
constant $\Lambda$ in this equation).
This renormalon is different from the familiar renormalon at $u=1/2$, 
which induces an $r$-independent uncertainty. (See footnote \ref{fn:10}.)
We note that this unfamiliar renormalon at $u=1/2$ can be an obstacle in 
estimating the normalization constants of the familiar renormalons at $u=1/2$ and $3/2$. 
This possibility is taken into account later in numerical analyses, while
we also present naive analysis by simply neglecting this peculiar renormalon. 

We make comments on the $\Delta V$-dependence of $V_S(r)$.
As already mentioned, perturbative computations of $V_S(r)$ do not
include $\Delta V$ as an external scale.
However, $\Delta V$ appears (only) in an implicit way
in a form $\alpha_s \times (\text{soft scale})$. 
Hence, the above renormalon uncertainty of $\mathcal{O}(\LMS r^2 \Delta V^2)$ in $\delta E_{\rm US}$
corresponds to the renormalon uncertainty in $V_S(r)$ of $\mathcal{O}(\LMS \alpha_s(1/r)^2)$,
where $\Delta V$ is expanded in $\alpha_s$.

One might then wonder if such implicit $\Delta V$-dependence 
ruins the argument in Sec.~\ref{sec:3.3} that the $u=3/2$ renormalon 
uncertainty of $V_S(r)$ is independent of $\Delta V$.
In fact this argument is correct for the $u=3/2$ renormalon;
the $u=3/2$ renormalon of $V_S(r)$ is canceled against the
leading UV contribution
of $\delta E_{\rm US}$, where $\Delta V$ in Eq.~\eqref{deltaEUS-energyintegral} is not relevant.

\medbreak\noindent
\underline{Scheme (B)}
\medbreak
We subtract IR divergences from $V_S(r)$ by adding
$\delta E_{\rm US}$ evaluated in double expansion in
$\alpha_s$ and $\log\alpha_s$
(Scheme B1). In this scheme, we do not distinguish $V_S(r)$ and $\delta E_{\rm US}$,
and we exclusively treat the sum of them, which is regarded as a redefined $V_S(r)$.
In this way we can subtract the IR divergences.
Furthermore, after canceling the IR divergences,
we can replace the argument of US logarithms
as $\log (r\Delta V) \to \log (r \mu_f)$
(Scheme B2).
Since both $\Vpos$ and
$\Delta V(r)$ are  $\mu$ independent,
the renormalized $V_S(r)$'s are also $\mu$ independent 
up to ${\cal O}(r^2)$
(although $V_S^{\rm (B2)}(r)$ is $\mu_f$ dependent).
\medbreak

Finally we point out that it is not straightforward to cancel simultaneously
both the  IR divergences and IR renormalon at $u=3/2$ of $V_S$ in the sum $V_S+\delta E_{\rm US}$.
We can observe this in the large-$\beta_0$ approximation. 
The renormalon uncertainty of $\delta E_{\rm US}$ coincides with minus that of $V_S(r)$
when $\Delta V$ in $\delta E_{\rm US}$ is not expanded in $\alpha_s$.
 If $\Delta V$ is perturbatively expanded instead, 
the power of $\alpha_s$ shifts by three in the perturbative series due to $\Delta V=\mathcal{O}(\alpha_s)$, 
as seen from eq.~\eqref{deltaEUSlargebeta0},
and the renormalon cancellation breaks down.\fn{
The normalization of the renormalon is changed by the expansion,
which ruins the renormalon cancellation.}
Hence, it would be optimal not to expand $\Delta V$ in $\alpha_s$ for the renormalon cancellation.
On the other hand, this prescription is not preferable to cancel the IR divergences.
The IR divergences are cancelled when $\Delta V$ is expanded in $\alpha_s$
as the IR and UV divergences in $V_S$ and $\delta E_{\rm US}$, respectively, appear at $\mathcal{O}(\alpha_s^4)$.
The proposed two schemes above can remove the IR divergences from $V_S(r)$,
but cannot remove the IR renormalons of $V_S(r)$. 
It remains a future task to develop a method for
subtracting the IR renormalons completely.\fn{
Suppose that we can remove completely
the $u=1/2$ and $3/2$ IR renormalons from $V_S$
 defined in the scheme B in some way.
The remaining renormalons are proportional to 
$(r^2\,\LQ^4/\Delta V)\times (\LQ/\Delta V)^n$ ($n\ge 0$)
or that with $\Delta V$ replaced by $\mu_f$.
They are obtained by expanding the correlator of
eq.\eqref{MatchingRel} in $t$.
In particular the leading IR renormalon ($n=0$) is given 
in terms of the local gluon condensate \cite{Voloshin:1978hc,Leutwyler:1980tn,Flory:1982qx,Brambilla:1999xf},
\bea
- i \, \frac{V_A(r)^2 }{6}\,\frac{r^2}{12\Delta V(r)}
\bra{0}g^2 G^{\mu\nu a}(0) G_{\mu\nu}^a(0)\ket{0}
\,,
\eea
or $\Delta V$ replaced by $\mu_f$.
Thus, the leading renormalon in $V_S^{\rm (B1,B2)}(r)$
is located at $u=2$ and suppressed by
$\LQ/\Delta V$ or $\LQ/\mu_f$ compared to the original $u=3/2$ renormalon in
$V_S(r)$.}

\subsection{Renormalon subtraction by contour deformation}

One motivation of the above investigation is to give
a justification to
the method used to subtract the $u=1/2$ and $3/2$ 
renormalons from $\Vpos$
in
a recent determination of $\alpha_s(M_Z)$ \cite{Takaura:2018lpw}.
There, it is assumed that the corresponding renormalons contained in
$\Vmos$ can be neglected.
(The IR divergences are canceled in momentum space.)
Then, using the one parameter integral form with respect to the
momentum $q$ and deforming the integral contour in the complex
$q$-plane, the renormalons at $u=1/2$ and $3/2$ which
stem from the original $q$ integral are subtracted.
See Refs.~\cite{Sumino:2005cq,Takaura:2018lpw} for the details.
As we have seen above,
the normalization of
the $u=1/2$ renormalon in the momentum-space potential is
exactly zero, while
the $u=3/2$ renormalon is suppressed by $\alpha_s(q)^2$.
While the $r^2\LMS^3$ renormalon that is generated purely
by the $q$ integral is subtracted, the suppressed renormalon in 
$\tilde{\alpha}_V(q)$ can still contribute to the position-space potential.
That is, if $\tilde{\alpha}_V(q)$ exhibits renormalon divergence, its uncertainty
will give an uncertainty to the renormalon-subtracted 
$V_S(r)$ 
constructed by
the contour deformation method.
It is expected to generate
a renormalon of order
$ r^2 \LMS^3 \alpha_s(1/r)$
in the renormalon-subtracted $V_S(r)$, corresponding
to the correction proportional to $\tilde{c}_1$ of Sec.~3.3.

\section{\boldmath Numerical study of $u=1/2$ renormalon}
\label{sec:4}

In the rest of this paper we perform numerical analyses on the normalizations 
$N_u$ of renormalons to check the above observations and to see the current status
of our knowledge on the perturbative series for $V_S(r)$ and $\Vmos$.
We treat two perturbative series:
one is the fixed order perturbative series, and the second one
includes higher-order terms estimated by RG,
which is used extensively in Refs.~\cite{Sumino:2003yp,Sumino:2005cq}.

In the case that the renormalon singularity of 
the Borel transform is given by eq.~\eqref{AnalyFormBX},
we can estimate the normalization constant $N_{u}$ from the fixed-order result
of the perturbative series.
This was first proposed in Ref.~\cite{Lee:1996yk}, whose method is as follows.
We consider the function
\be
N(t)=\lt(1-\frac{b_0 t}{u} \rt)^{1+\nu_u} B_X(t;\mu=Q) \, .
\label{NLee}
\ee
(In the following analyses, $\nu_u$ of eqs.~\eqref{nuonehalf} and \eqref{nuthreehalf} are used.)
We can obtain the normalization constant by expanding this function in $t$ and then substituting $t=u/ b_0$:
\be
N_u=\sum_{i=0}^{\infty} N_i t^i |_{t=u/b_0} \, ,
\label{expNu}
\ee
as long as the corresponding renormalon is the closest one 
to the origin.\footnote{
Note that the regular part of $B_X$ at $t=u/b_0$ would generate, e.g.,
the series expansion of $(1-{b_0 t}/{u} )^{1+\nu_u}$ in $t$ which
is convergent at $t= u/b_0$ (even though it is divergent at 
$|t|>u/b_0$).
}
This method is fairly general and can be used with only known terms of the
perturbative series.

Using this method,  Ref.~\cite{Pineda:2001zq}  first studied
$N_{1/2}$ from the fixed-order result.  
In Ref.~\cite{Ayala:2014yxa}, $N_{1/2}$ is estimated 
with a more recent NNNLO result \cite{Anzai:2009tm,Smirnov:2009fh}.
We present the NNNLO result \cite{Lee:2016cgz} by  
subtracting the IR divergence at NNNLO in the scheme (A),
\begin{align}
N_{1/2}
&=-1.3333 \quad \text{at LO} \non
&=-0.76139 \quad \text{at NLO} \non
&=-1.10661 \quad \text{at NNLO} \non
&=-1.21655 \quad \text{at NNNLO} \, .
\end{align}
[If we adopt the scheme (B2), the NNNLO result is given by 
$N_{1/2}=-1.07764-0.0541542 \log(2 r \mu_f)$.]
We note that the IR logarithm in the NNNLO perturbative coefficient
considered here is different from Ref.~\cite{Ayala:2014yxa}.
We adopt the result in dimensional regularization given in ref.~\cite{Anzai:2009tm}
(more precisely its Fourier transform), 
while ref.~\cite{Ayala:2014yxa} adopted the result in Ref.~\cite{Brambilla:1999qa}.
(As a result, the IR logarithms differ by factor $4$.)
This difference corresponds to different choices of renormalization scheme.

\begin{figure}[t!]
\begin{center}
\includegraphics[width=9cm]{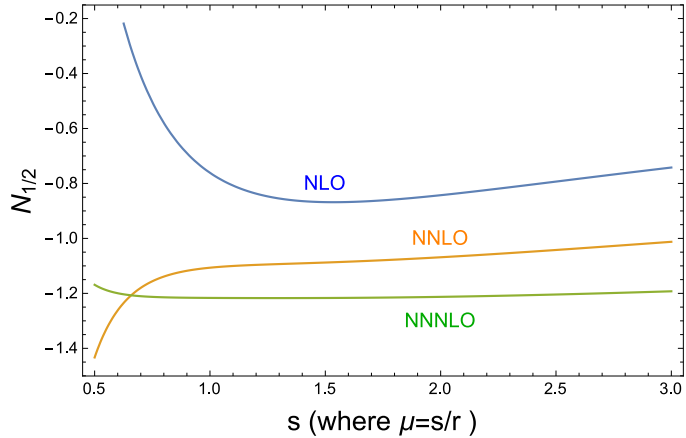}
\end{center}
\caption{\small
Scale dependence of the normalization constant $N_{1/2}$.}
\label{fig:scaledeponehalf}
\end{figure}

We also examine the scale dependence of the estimated normalization constant.
We use the perturbative coefficients with the renormalization scale $\mu=s/r$ to estimate the normalization constant.\fn{
When the scale $\mu=s/r$ is used in constructing the Borel transform,
the normalization constant $N_u(s)$ of the renormalon at $u$ behaves as $N_u(s)=N_u(s=1) s^u$
as seen from eq.~\eqref{AnalyFormBX}. The $s$-dependence of the estimated result for $s^{-u} N_u(s)$ 
is expected to reduce as we include higher-order terms.
We always consider $N_u(s=1)$ unless stated otherwise explicitly. 
}
The results are shown in Fig.~\ref{fig:scaledeponehalf}.
The scale dependence decreases as we include higher-order terms.
These results indicate that the series \eqref{expNu} shows convergence
for the $u=1/2$ renormalon and $N_{1/2}(s=1)\simeq -1.1$.


In the second method we treat the RG-improved series,
where the higher-order terms in perturbative series are estimated by RG.
Explicitly we can use~\cite{Sumino:2003yp,Sumino:2005cq}
\be
v(r)=-\frac{2 C_F}{\pi} \int_0^{\infty} \frac{dq}{q} \, \sin(q r) [\widetilde{\alpha}_V(q)]_{\text{N$^k$LL}} \, , \label{pot2}
\ee
where the N$^k$LL terms of the perturbative series
[coefficients of $\alpha_s(\mu)^{n+k+1}\,\log^n(\mu r)$ for
arbitrary $n$]
can be determined using the $(k+1)$-loop beta function
and the fixed-order result up to $k$-loops.

We now estimate $N_{1/2}$ from the RG-improved series obtained from eq.~\eqref{pot2}
using eq.~\eqref{expNu}.
Since we have an all-order perturbative series (at each order of improvement),
we can obtain $N_{\rm LL},\cdots,N_{\rm N^3LL}$ with arbitrary precision in principle.
The results from finite number of terms read
\begin{align}
N_{1/2}
&=-0.8488 \quad \text{at LL  (using 30 terms)} \non
&=-0.7603 \quad \text{at NLL (using 30 terms)} \non
&=-1.009 \quad \text{at NNLL (using 30 terms)} \non
&=-1.120 \quad \text{at NNNLL (using 30 terms)}  \, . \label{est2ndfo}
\end{align}
We use the scheme (A) to subtract the IR divergence in the NNNLL analysis.\fn{
We perform Fourier transform of the finite result $V_S(r)$ obtained in the scheme (A)
to obtain regularized $\tilde{\alpha}_V(q)$ in momentum space. 
This is not equivalent to the regularization where we 
set $\mu=q$ in the NNNLO result of $\tilde{\alpha}_V(q)$ and subtract the $1/\epsilon$ term.
}
From Table~\ref{tab1}, which shows the convergence speed,
we infer that 20-50 perturbative coefficients 
are needed in order to obtain the normalization constants with 
one-percent accuracy.
\begin{table}[t]
\begin{center}
\begin{tabular}{c|D{.}{.}{6}D{.}{.}{6}D{.}{.}{6}D{.}{.}{6}}
\hline
\multicolumn{1}{c|}{The number of terms} & \multicolumn{1}{c}{LL} & \multicolumn{1}{c}{NLL} & \multicolumn{1}{c}{NNLL} & \multicolumn{1}{c}{NNNLL} \\ \hline
1 & -1.33333 & -1.33333    & -1.33333  &   -1.33333\\
2 & -0.769621 & -0.761390   & -0.761390 &   -0.761390\\
3 & -0.770430  & -0.553531  & -1.10661 &  -1.10661\\
4 & -0.893478 & -0.764315 &  -0.893382 & -1.21655\\
5 & -0.834305 & -0.777275 &  -0.973020 &   -1.03927\\
10 & -0.848814 & -0.758382 & -1.00381 &  -1.11924\\
15 & -0.848826 & -0.759417 & -1.00658 &   -1.11968\\
20 & -0.848826  & -0.759892 & -1.00779 &  -1.11995\\
25 & -0.848826   & -0.760151 & -1.00844 & -1.12010\\
\hline
\end{tabular}
\end{center}
\caption{\small
Estimates of normalization constant $N_{1/2}$ from truncated perturbative series of 
the RG-improved series.
\label{tab1}
}
\end{table}


\section{Renormalon cancellation in total energy}

It is interesting to examine renormalon cancellation in the total energy (namely, $V(r)+2 m_{\rm pole}$) 
from the estimated $N_{1/2}$ \cite{Pineda:2001zq}.
The leading renormalon in the Borel transform of $m_{\rm pole}/m_{\rm{\overline{MS}}}$ is 
given by
\be
B_{m_{\rm pole}/m_{\MSbar}}(\mu=m_{\MSbar}) 
\simeq \frac{N_M}{(1-2 b_0 t)^{1+\frac{b_1}{2 b_0^2}}} \, .
\ee
$N_M$ can be investigated from the fixed-order perturbative series \cite{Melnikov:2000qh,Marquard:2016dcn} in a parallel manner.
The results of $N_M$ are given by\footnote{
Note that the perturbative series needs to be expressed in terms of
the coupling of the theory with $n_l$ light quarks only, while originally
the pole mass is expressed by the coupling in the theory with $n_l$ light quarks plus one heavy quark.
This is needed to ensure the renormalon cancellation, since
$N_M$ and $N_{1/2}$ are proportional to  $\LMS$ and the same $\LMS$ 
should be used for both quantities.
(In principle one can
pursue the calculation in the different couplings if the difference
in the definitions of $\LMS$ is properly taken into account.)
} 
($\mu=m_{\rm{\overline{MS}}}$)
\begin{align}
N_M
&=0.424413 \quad \text{at LO} \non
&=0.562265 \quad \text{at NLO} \non
&=0.574979 \quad \text{at NNLO}  \non
&=0.513427 \pm 0.001025 \quad \text{at NNNLO}  \, .
\end{align}
The last result has an error due to the numerical error of the $\mathcal{O}(\alpha_s^4)$ coefficient.
Now let us examine the renormalon cancellation, $2 N_M+N_{1/2}=0$.
\begin{align}
\frac{2 N_M+N_{1/2}}{(2 N_M-N_{1/2})/2}
&=-0.444 \quad \text{at LO} \non
&=0.385 \quad \text{at NLO} \non
&=0.038 \quad {\text{at NNLO}} \non
&=-0.088\pm0.002 \quad {\text{at NNNLO}}
\end{align}

\begin{figure}[h]
\begin{center}
\includegraphics[width=9cm]{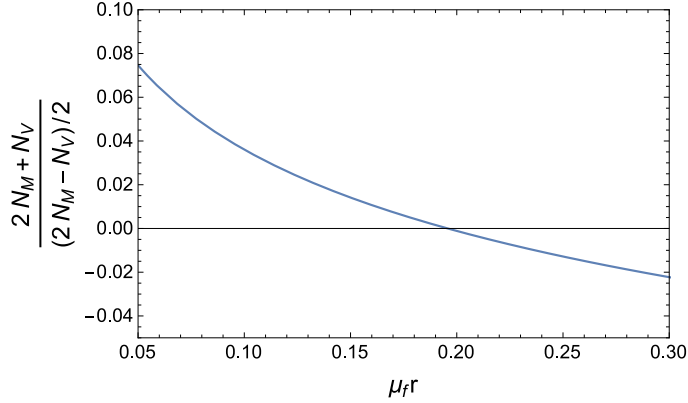}
\end{center}
\caption{\small
Renormalon cancellation as a function of $\mu_f$ in the logarithm.}
\label{fig1}
\end{figure}

It is possible that treatment of IR divergences affects the cancellation. 
Let us examine this.
So far, we subtracted the IR divergence in the scheme (A),
but now we make the three-loop coefficient finite in the scheme (B2).
Then, the IR divergence is replaced by the logarithmic term like 
$\log(\mu_f r)$.
In Fig.~\ref{fig1}, we investigate renormalon cancellation 
while varying $\mu_f$ in this logarithm in a reasonable range.
This figure shows that the treatment of IR divergences can be non-negligible 
to the precise cancellation.
We note that the numerical error on the four-loop result of the mass relation hardly affects this result.

\section{Normalization by analytic formula in RG-improved series}
\begin{figure}[t!]
\begin{minipage}{0.5\hsize}
\begin{center}
\includegraphics[width=6cm]{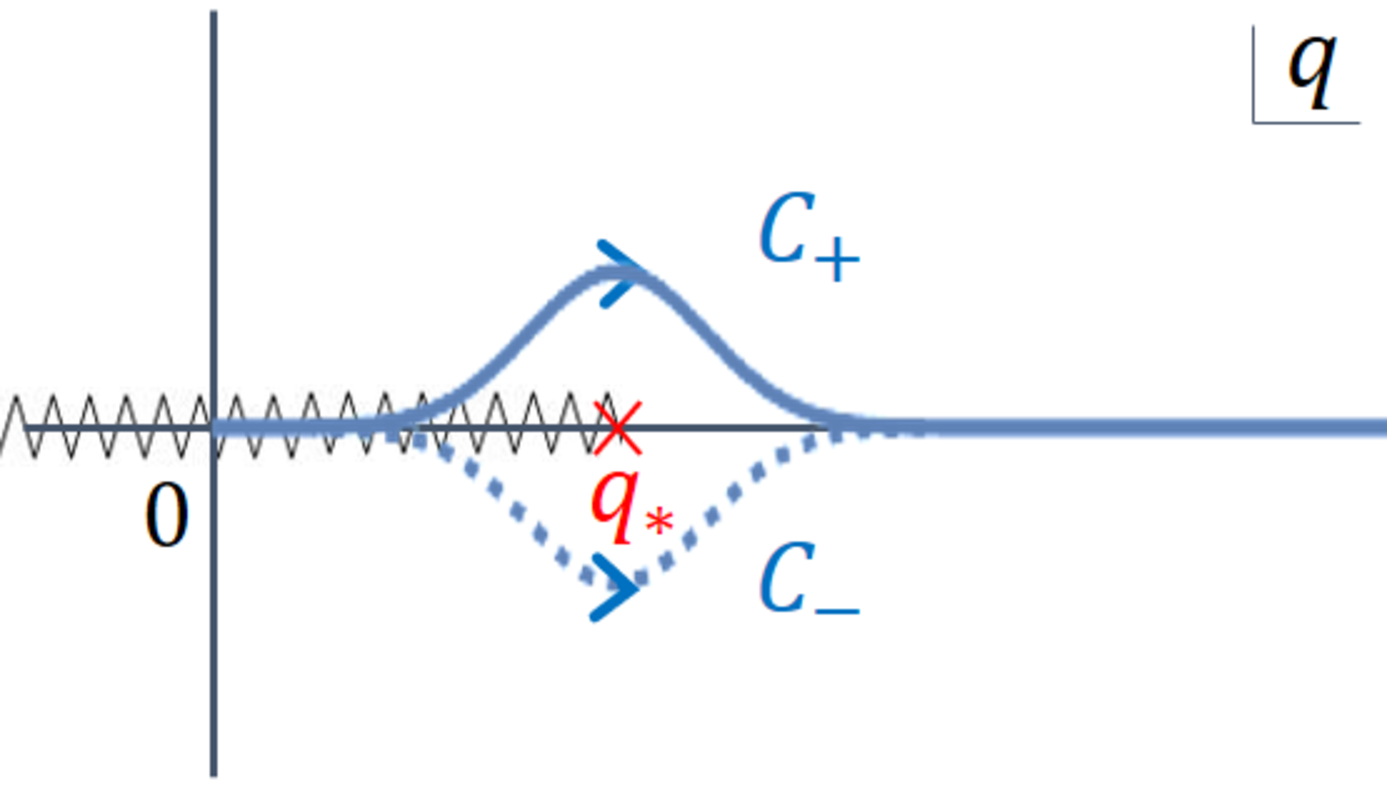}
\end{center}
\end{minipage}
\begin{minipage}{0.5\hsize}
\begin{center}
\includegraphics[width=6cm]{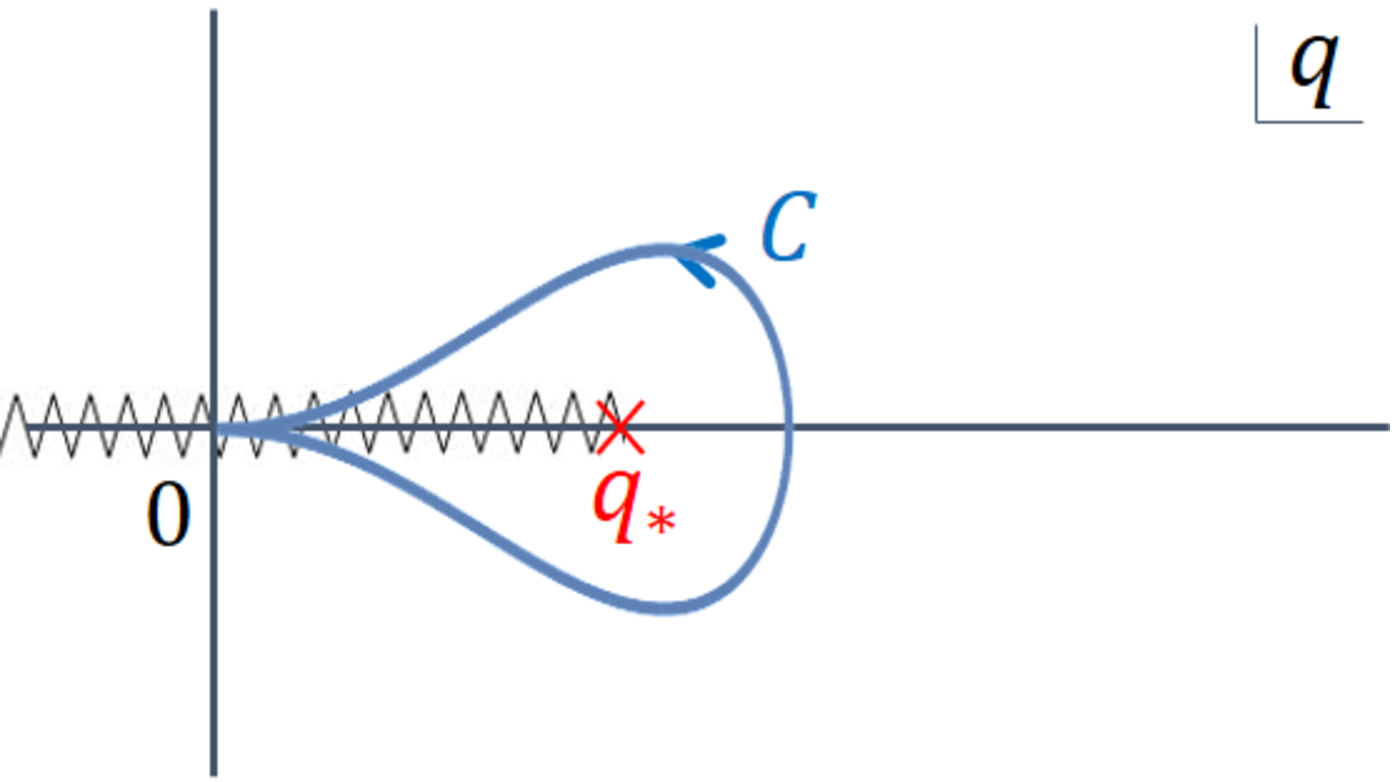}
\end{center}
\end{minipage}
\caption{\small
Contours $C_\pm$ and $C$. 
$q_*$ is the singular point of $\widetilde{\alpha}_V(q)$ on the positive real $q$ axis.
\label{Fig:C}
}
\end{figure}

For the RG-improved series, we derive a formula for the normalization constants of renormalons
given as a one-dimensional integral.
The Borel integral of the QCD potential for the RG-improved series 
can be written as
\be
\,{v}_{\pm}=-\frac{2 C_F}{\pi} \int_{C_\mp} \frac{dq}{q} \, \sin(q r) [\widetilde{\alpha}_V(q)]_{\text{N$^k$LL}} \, , 
\label{resum2}
\ee
where the contours $C_\mp$ are displayed in Fig.~\ref{Fig:C}.
The details of the derivation are given in App.~B.
Since the integrand satisfies $\{f(x) \}^*=f(x^*)$, the imaginary part 
can be calculated by a contour integral
\be
{\rm Im} \,{v}_{\pm}
=\mp \frac{2 C_F}{\pi} \frac{1}{2 i} \int_{C} \frac{d q}{q} \sin (qr) \widetilde{\alpha}_V(q) \, ,
\ee
where the contour $C$ is displayed in Fig.~\ref{Fig:C}.
By expanding $\sin (q r)$ in $q r$, the normalization
of the renormalon at $u=1/2$ is found as
\begin{align}
{\rm Im} \,{v}_{\pm}[{\textstyle u=\frac{1}{2}}]
&=\mp\frac{2 C_F}{\pi} \frac{1}{2 i} r \int_C dq \,  \widetilde{\alpha}_V(q)  \non
&=\mp \frac{2 C_F}{\pi} \LMS r \frac{1}{2 i}  \int_C dx \,  
\widetilde{\alpha}_V(x) \label{amb2} \,.
\end{align}
In the last line, we changed the integration variable to $x=q/\LMS$.
(Note that $\widetilde{\alpha}_V(q)$ is a function of $q/\LMS$.)
Then, from eqs.~\eqref{ambX} and \eqref{amb2}, we obtain 
\be
N_{1/2}=-\frac{2 C_F}{\pi} \frac{b_0}{\pi} \Gamma(1+u b_1/b_0^2) u^{-1-u \frac{b_1}{b_0^2}} I_{u}|_{u=1/2} \, , \label{ana}
\ee
with
\be
I_{1/2}=\frac{1}{2 i} \int_C d x \, \widetilde{\alpha}_V(x) \, .
\ee

We present numerical values of $N_{1/2}$ via numerical evaluation of $I_{1/2}$:
\begin{align}
N_{1/2}
&=-0.848826 \quad \text{at LL} \non
&=-0.760846 \quad \text{at NLL} \non
&=-1.01017 \quad \text{at NNLL} \non
&=-1.12049 \quad \text{at NNNLL}  \, .
\end{align}
They agree well with the estimates from the finite number of terms \eqref{est2ndfo}.
The scheme (A) is adopted at NNNLL in accordance with Sec.~\ref{sec:4}.

It is possible to calculate the normalization constants of other renormalons in a parallel manner. The normalization constant of a general renormalon at $u$ is expressed as
\be
N_{u}=-\frac{2 C_F}{\pi} \frac{b_0}{\pi} \Gamma(1+u b_1/b_0^2) u^{-1-u \frac{b_1}{b_0^2}} I_u \label{generalN}
\ee
with
\be
I_u=\frac{1}{2 i} \int_C dx \, \frac{(-1)^{(2 u-1)/2}}{(2u)!} x^{2 u-1}  \widetilde{\alpha}_V(x) \, . \label{generalI}
\ee
This expression stems from the Taylor expansion of $\sin (qr)$.

In this method, the $r$-dependence of a renormalon uncertainty 
due to any half-integer renormalon at $u$ is given exactly by $r^{2u+1}$.
In particular for $u=3/2$, the correction factor of $[1+\mathcal{O}(\alpha_s(1/r))]$
is not detected (as long as we work at N$^k$LL with finite $k$).
It is because this method relies on
the assumption that $\alpha_V(q)$ does not possess renormalon uncertainties.

For the $u=3/2$ renormalon, which is the second IR renormalon, 
the numerical values of $N_{3/2}$ are given by
\begin{align}
N_{3/2}
&=0.0471570 \quad \text{at LL} \non
&=0.0260142 \quad \text{at NLL} \non
&=0.0793089 \quad \text{at NNLL} \non
&=0.143286 \quad \text{at NNNLL}  \, , \label{ana3/2}
\end{align}
based on eqs.~\eqref{generalN} and \eqref{generalI}.
We again adopt the scheme (A).

\section{\boldmath Numerical analysis of $u=3/2$ renormalon}

We now estimate $N_{3/2}$ from the fixed-order result.
We annihilate the leading renormalon at $u=1/2$, whose uncertainty
is an $r$-independent constant, by considering the QCD force.
Then we use the same method as in the $u=1/2$ renormalon.

We first examine the relation between 
the normalization constants of the potential and force.
The potential $v=r V$ has the $u=3/2$ renormalon uncertainty as eq.~\eqref{ambX} with eqs.~\eqref{nuthreehalf} 
and \eqref{c1tildethreehalf}, 
which gives the uncertainty to the dimensionless force $f=r^2 dV/dr$ as
\be
{\rm Im} \,{f}_{\pm}=\pm  \frac{\pi}{b_0} \frac{2 N_{3/2}}{\Gamma(1+u b_1/b_0^2)}  u^{1+u \frac{b_1}{b_0^2}} (r^2 \LMS^2)^{3/2} [1+\mathcal{O}(\alpha_s(1/r))] |_{u=3/2} \, .
\ee 
Thus, the normalization constant of the dimensionless force $N_{3/2}^F$ is related as $N_{3/2}=\frac{1}{2} N_{3/2}^F$.

To obtain $N_{3/2}$,
we first consider the fixed-order perturbative series of the potential 
without setting $\mu=1/r$. 
The derivative with respect to $r$ gives the fixed-order result of the force. 
Finally we set $\mu=1/r$ to estimate $N_{3/2}^F$ and translate it to $N_{3/2}$.

\begin{figure}[tbp]
\begin{center}
\includegraphics[width=9cm]{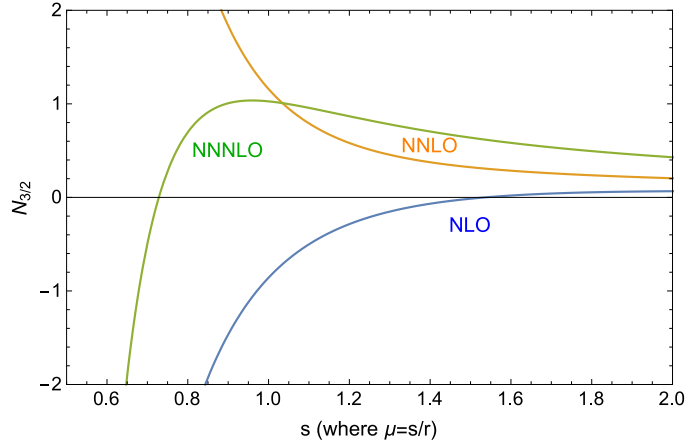}
\end{center}
\caption{\small
Scale dependence of the normalization constant $N_{3/2}$.}
\label{fig:scaledepthreehalf}
\end{figure}

We present the results:
\begin{align}
N_{3/2}
&=0.666667 \quad \text{at LO} \non
&=-0.857914 \quad \text{at NLO} \non
&=1.15844 \quad \text{at NNLO} \non
&=1.02659 \quad \text{at NNNLO}  \, .
\end{align}
We adopt the scheme (A) to obtain the NNNLO result.
[In the scheme (B2), we obtain $N_{3/2}=-0.848623+0.731082 \log(2 \mu_f r)$ at NNNLO.]
Although the estimate of the normalization constant may look already convergent,
this seems to be a numerical accident.
We examine the scale dependence of the estimated normalization constant in a parallel manner to the $u=1/2$ renormalon.
The result is shown in Fig.~\ref{fig:scaledepthreehalf}.
We find that  large dependence on the renormalization scale remains in this estimate.

Let us perform a parallel estimate from the finite order result 
of the RG-improved series.
In this case, since we know $N_{3/2}$ as given by 
eq.~\eqref{ana3/2}, it would be useful to grasp 
how many terms are needed for a reasonable estimate.
Table~\ref{tab2} shows the result. At NNNLL, we adopt the scheme (A).
One sees that typically 20 terms are needed for a good estimate.
\begin{table}[t!]
\begin{center}
\begin{tabular}{c|D{.}{.}{7}D{.}{.}{7}D{.}{.}{7}D{.}{.}{7}}
\hline
\multicolumn{1}{c|}{The number of terms} & \multicolumn{1}{c}{LL} & \multicolumn{1}{c}{NLL} & \multicolumn{1}{c}{NNLL} & \multicolumn{1}{c}{NNNLL} \\ \hline
1 & 0.666667 & 0.666667    & 0.666667 &  0.666667\\
2 & -0.845569 &-0.857914    & -0.857914 &    -0.857914\\
3 & 0.00364088 & -1.33043  & 1.15844 &    1.15844\\
4 & 1.66115 &3.05043 &  -3.33613 &   1.02659\\
5 & -2.39650 & 0.00590101 &  3.26898 & -7.65424\\
10 & -0.525249 & 1.23022 & -1.32332 &   -5.59838\\
15 & 0.0300462 & -0.00224226 & 0.1238239 & 0.294562\\
20 & 0.0471086 & 0.0254835 & 0.0802829 & 0.151007\\
25 & 0.0471576 &0.0260084 & 0.0791596 & 0.143535\\
30 & 0.0471570 & 0.0260113 & 0.0792097 & 0.143424\\
\hline
\end{tabular}
\end{center}
\caption{\small
Estimates of the normalization constant 
$N_{3/2}$ from truncated perturbative series of the RG-improved series.
}
\label{tab2}
\end{table}

\begin{figure}[t!]
\begin{minipage}{0.5\hsize}
\begin{center}
\includegraphics[width=7cm]{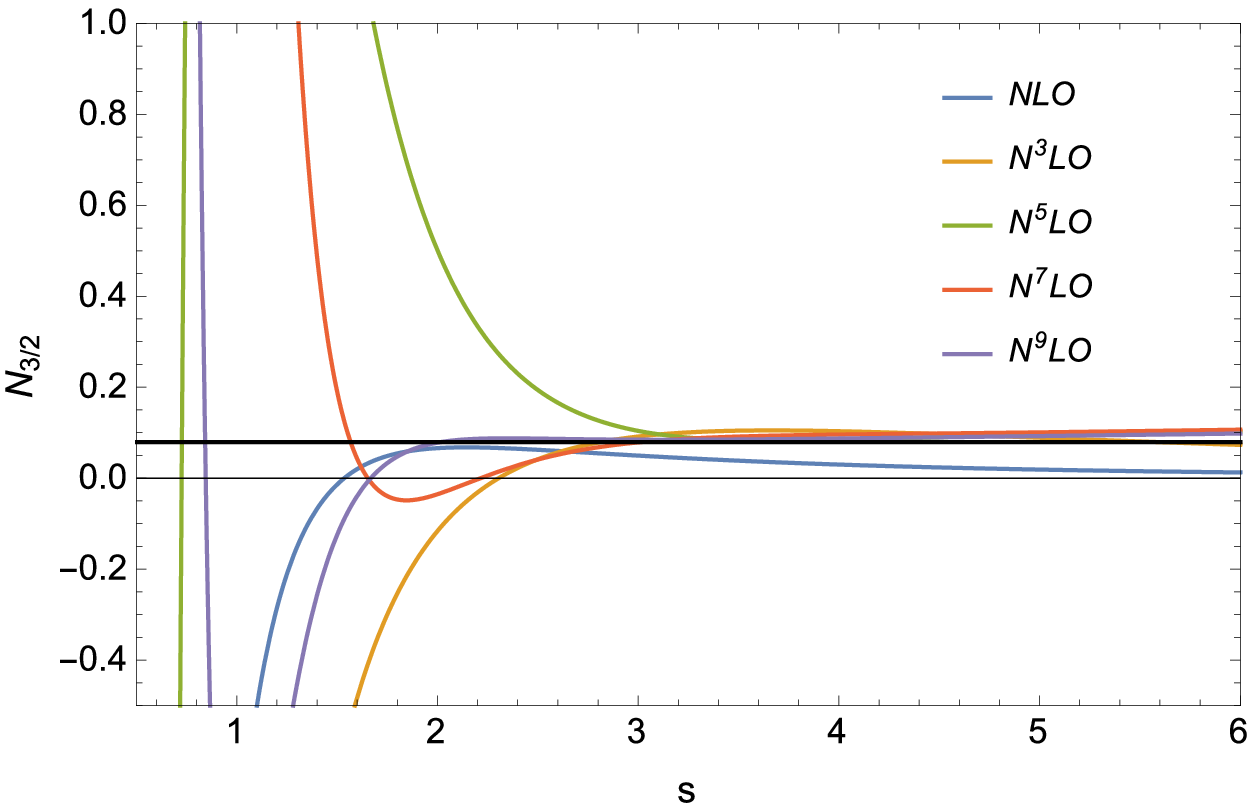}
\end{center}
\end{minipage}
\begin{minipage}{0.5\hsize}
\begin{center}
\includegraphics[width=7cm]{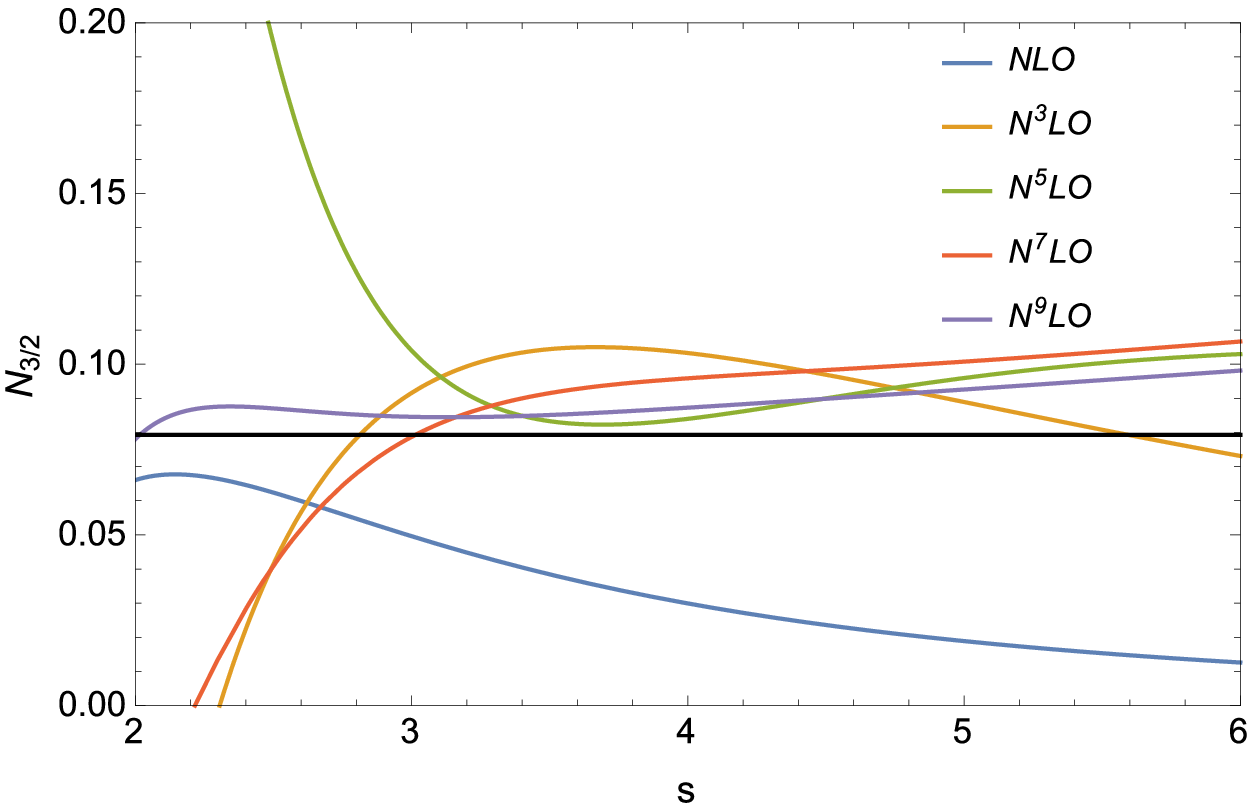}
\end{center}
\end{minipage}
\caption{\small
Scale dependence of $N_{3/2}$ determined from finite number of terms at NNLL.
Black line shows the exact answer obtained from the analytic formula.
In the right panel, the range $2 \leq s \leq 6$ is magnified.}
\label{fig:ScaledepinRGimp}
\end{figure}

We examine scale dependence of the 
estimate of $N_{3/2}$ using finite number of terms in the RG-improved series.
Since we know the exact answer in this case, we can directly check whether 
mild scale dependence indicates reliability of the estimate.
In Fig.~\ref{fig:ScaledepinRGimp}, we examine this at NNLL. 
We see that at higher order the scale dependence of $N_{3/2}$
decreases
and it approaches the correct value.
(At further higher order, for instance from 30 terms, we obtain $N_{3/2}=0.0079136$ and $0.079172$ for 
$s=1/2$ and $2$, respectively.)

In Sec.~\ref{sec:3.5}, we pointed out that $V_S(r)$ can have the unfamiliar renormalon at $u=1/2$
associated with the IR divergences. 
Since the corresponding renormalon uncertainty for the static QCD potential
is not an $r$-independent constant in contrast to the familiar renormalon at $u=1/2$,
this renormalon cannot be eliminated in the QCD force.
Taking into account this possibility, 
we present another estimate for the $u=3/2$ renormalon,
whose method is not plagued by the two renormalons at $u=1/2$.
We carry out this by using a mapping from the $t$-plane to a new $z$-plane,
where the $u=3/2$ renormalon becomes closer to the origin than the $u=1/2$ renormalon.
Namely we change the relative distances of the two IR renormalons from the origin.\footnote{
One may compare with
Ref.~\cite{Lee:1996yk}, in which 
 for the Adler function
the closer UV renormalon at $u=-1$
is made farther than the IR renormalon at $u=2$ by a mapping.
}

A possible mapping is given by 
\be
b_0 t (z)=\frac{1}{2} (z+e^{i \pi/6})^6+\frac{1}{2} \, . \label{map1}
\ee
The basic idea to obtain this mapping is as follows.
In the first step, we consider $v=2 (b_0 t)-1$,
which maps $b_0 t=0$, $1/2$, $3/2$ into $v=-1$, $0$, $2$, respectively.
In the second step, $w=v^{1/6}$ is considered,
which makes the distance between $v=-1$ $(u=0)$ and $v=2$ $(u=3/2)$ shorter
than that between $v=-1 (u=0)$ and $v=0 (u=1/2)$.
The final step is given by $z=v-e^{i \pi/6}$ to locate the original origin $u=0$ at $z=0$.
Corresponding to these transformations, we consider Eq.~\eqref{map1}. 
Indeed, the closest zero of $1- b_0 t (z)/u $ among positive half-integers $u$ 
is given by $u=3/2$.

However, it turned out that  with the above mapping convergence is too slow for practical analysis.
(e.g.,
In RG-improved series at LL, we need 250 terms to obtain the normalization constant 
with about 10 \% accuracy.)
Instead of Eq.~\eqref{map1}, we use
\be
b_0 t(z)=2-\frac{1}{2} [9-7 (e^{i \pi/6}-z)^6]^{1/2} \, . \label{map2}
\ee
This mapping is obtained with a similar idea to the above,
but the main difference is that we first consider square of the difference from $b_0 t=2$,
i.e. $(2-b_0 t)^2$.  
The mapping \eqref{map2} consists of the following steps: $ b_0 t (v)=2-v^{1/2}$, 
$v(w)=-\frac{7}{4} w+\frac{9}{4}$, $w(y)=y^6$, $y(z)=-z+e^{i \pi/6}$.
We note that the singularities of $1/(1-2 b_0 t(z)/3)$ and $1/(1-2 b_0 t(z)/5)$
with respect to $z$ are not common, 
and the $u=5/2$ renormalon does not affect the estimate of the normalization 
constant at $u=3/2$.
With this mapping $t(z)$, we consider a function
\be
N(z)= \lt(1- \frac{2}{3} b_0 t (z) \rt)^{1+\frac{3}{2} \frac{b_1}{b_0^2} } B_v (t (z); \mu=1/r) \, .
\ee
By expanding this function in $z$ and then substituting $z=-(8/7)^{1/6}+e^{i \pi/6}$,
we can obtain the normalization constant of the $u=3/2$ renormalon.

Using this mapping, we estimate the normalization constant of the $u=3/2$ renormalon
from the fixed-order perturbative series as
\begin{align}
N_{3/2}
&=-1.33333 \quad \text{at LO} \non
&=-1.52383 \quad \text{at NLO} \non
&=4.75182 \quad \text{at NNLO} \non
&=9.01375 \quad \text{at NNNLO} \, .
\end{align}
We adopt the scheme (A) at NNNLO.
[In the scheme (B2), we obtain $N_{3/2}=7.2395+0.69179 \log(2 \mu_f r)$.]
In this method, imaginary parts appear in fixed-order results, but we
omit them in the above estimate since we know that the true normalization is real. 
The size of the imaginary parts can be used for an error estimate of the results.

We also estimate $N_{3/2}$ of the RG-improved series using this mapping.
Table~\ref{tab3} shows the result.
One can confirm that the estimated values converge to the results in Eq.~\eqref{ana3/2}.
We start to obtain reasonable results with about 60 terms.
\begin{table}[h!]
\begin{center}
\begin{tabular}{c|D{.}{.}{5}D{.}{.}{5}D{.}{.}{5}D{.}{.}{5}}
\hline
\multicolumn{1}{c|}{The number of terms} & \multicolumn{1}{c}{LL} & \multicolumn{1}{c}{NLL} & \multicolumn{1}{c}{NNLL} & \multicolumn{1}{c}{NNNLL} \\ \hline
1 &  -1.33333 &-1.33333    & -1.33333 &  -1.33333\\
2 & -1.52878 &-1.52383    & -1.52383 &    -1.52383\\
3 & 1.20631 & 0.96708  & 4.75182 &    4.75182\\
4 & 4.86579 & 6.76949 &  4.88592 &   9.01375\\
5 & 6.13184 & 6.39316 &  -9.64487 &   8.39204\\
10 & -1.32119 & 30.9456 & -220.002 &  277.649\\
20&  -12.5409 & 20.1377  & 75.0878 & -347.457\\
30 & 4.04035 & -14.7808 & 66.2768 & -32.0964\\
40 & -1.07937 & 0.943337 & -5.54367 & -70.4336\\
50 & 0.0725315 & 0.0420682 & 0.0878506 & 2.16533\\
60 & 0.047093 & 0.0258843 & 0.078969  & 0.167635 \\
100 & 0.0471979 & 0.0262156   & 0.0801885 & 0.143394 \\  
150 & 0.0471763 & 0.0260685 & 0.0794704 & 0.143429\\
200 & 0.0471475  & 0.0259836& 0.0792261 & 0.143089\\ \hline
\end{tabular}
\end{center}
\caption{\small
Estimates of the normalization constant 
$N_{3/2}$ with using mapping \eqref{map2}
from truncated perturbative series of the RG-improved series.
}
\label{tab3}
\end{table}

\newpage
\section{\boldmath $u=1/2$ and $3/2$ renormalons in $\Vmos$}

Let us investigate the renormalon uncertainty of $\widetilde{\alpha}_V(q)$.
We estimate the
normalization constants of the renormalons of $\widetilde{\alpha}_V(q)$ at $u=1/2$ and $3/2$ 
assuming that they are the leading renormalon individually.
More explicitly we assume
\be
B_{\widetilde{\alpha}_V(q)}(t) \simeq \frac{N_{\widetilde{\alpha}_V, u}}{(1-\frac{b_0}{u} t)^{1+u \frac{b_1}{b_0^2}}} \, .
\ee
The theoretical discussion in Sec.~\ref{sec:3.4} shows that the normalization constants $N_{\widetilde{\alpha}_V, u}$
defined in this way should be zero both for $u=1/2$ and $u=3/2$, because
the $u=1/2$ renormalon is completely absent in $\tilde{\alpha}_V(q)$, and for $u=3/2$
the expansion of the Borel transform around the singularity takes a form $\sim (1-b_0t/u)^{-1-u b_1/b_0^2+2}$
rather than $\sim (1-u  b_0t/u)^{-1-u b_1/b_0^2}$ corresponding to the $\alpha_s(q)^2$ suppression.\footnote{
As a result of the suppression of the renormalon, 
we may regard that $\nu_{3/2}$ for the momentum-space potential is shifted by $-2$.
}

The estimates from the fixed-order results read
\begin{align}
N_{\widetilde{\alpha}_V, u=1/2}
&=1 \quad \text{at LO} \non
&=-0.00617284 \quad \text{ at NLO} \non
&=0.141682  \quad \text{at NNLO} \non
&=-0.0318992  \quad \text{at NNNLO}
\end{align}
and
\begin{align}
N_{\widetilde{\alpha}_V, u=3/2}
&=1 \quad \text{at LO} \non
&=-0.0185185 \quad \text{at NLO} \non
&=2.246032  \quad \text{at NNLO} \non
&=2.27727   \quad \text{at NNNLO} \, .
\end{align}
We subtract the IR divergence at NNNLO in the scheme (A).
If we instead use the scheme (B2),
the NNNLO results are modified as
\be
N_{\widetilde{\alpha}_V, u=1/2}=-0.136078+0.0406157 \log({2 r \mu_f})  \quad \text{at NNNLO} \, ,
\ee
and
\be
N_{\widetilde{\alpha}_V, u=3/2}=-0.535547+1.09662 \log({2 r \mu_f})  \quad \text{at NNNLO} \, .
\ee
By taking $r \mu_f=0.2$ as an example, 
we obtain $N_{\widetilde{\alpha}_V, u=1/2}=-0.1723293$
and $N_{\widetilde{\alpha}_V, u=3/2}=-1.54037$ at NNNLO.

\begin{figure}[t!]
\begin{minipage}{0.5\hsize}
\begin{center}
\includegraphics[width=7cm]{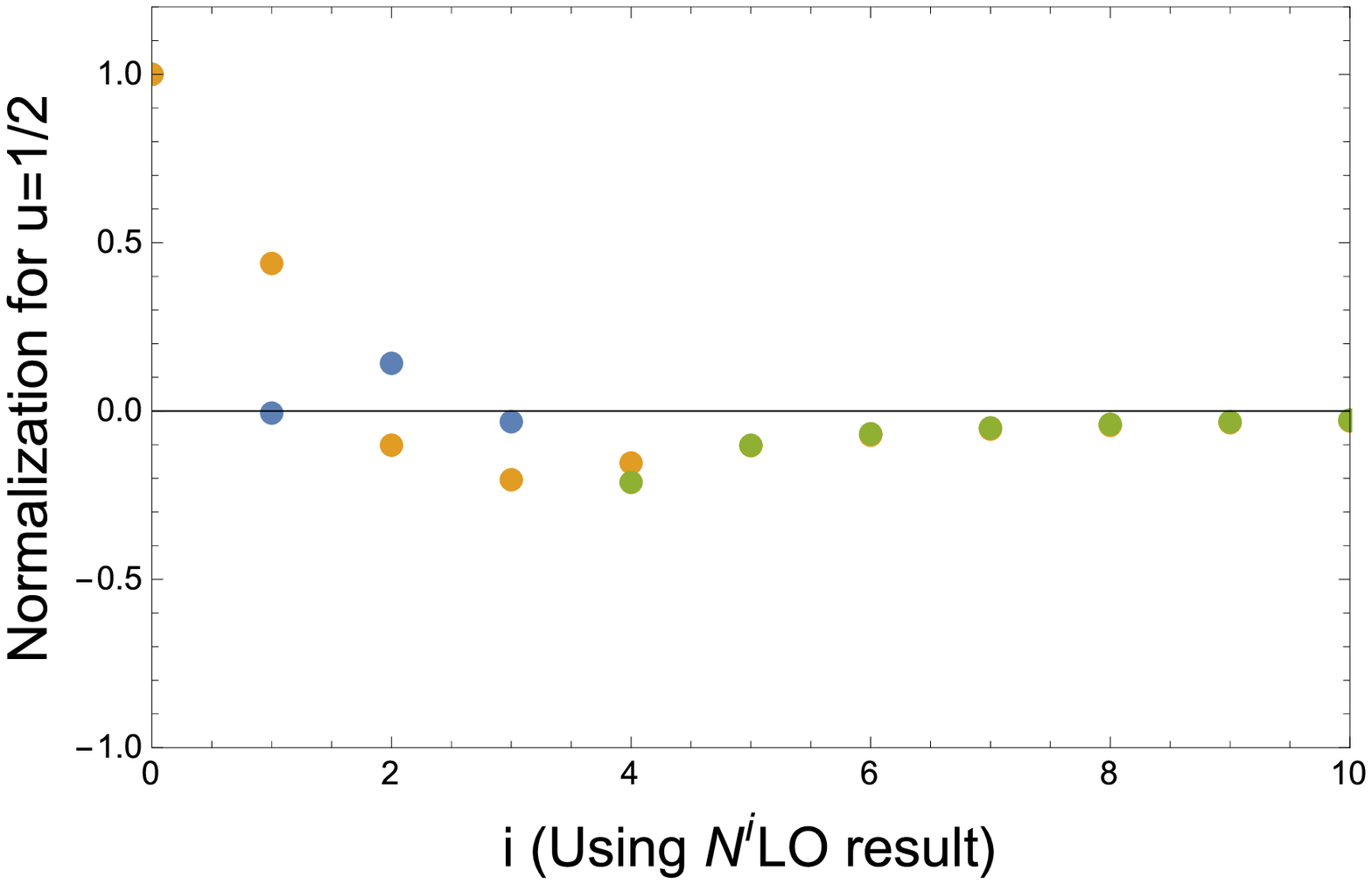}
\end{center}
\end{minipage}
\begin{minipage}{0.5\hsize}
\begin{center}
\includegraphics[width=7cm]{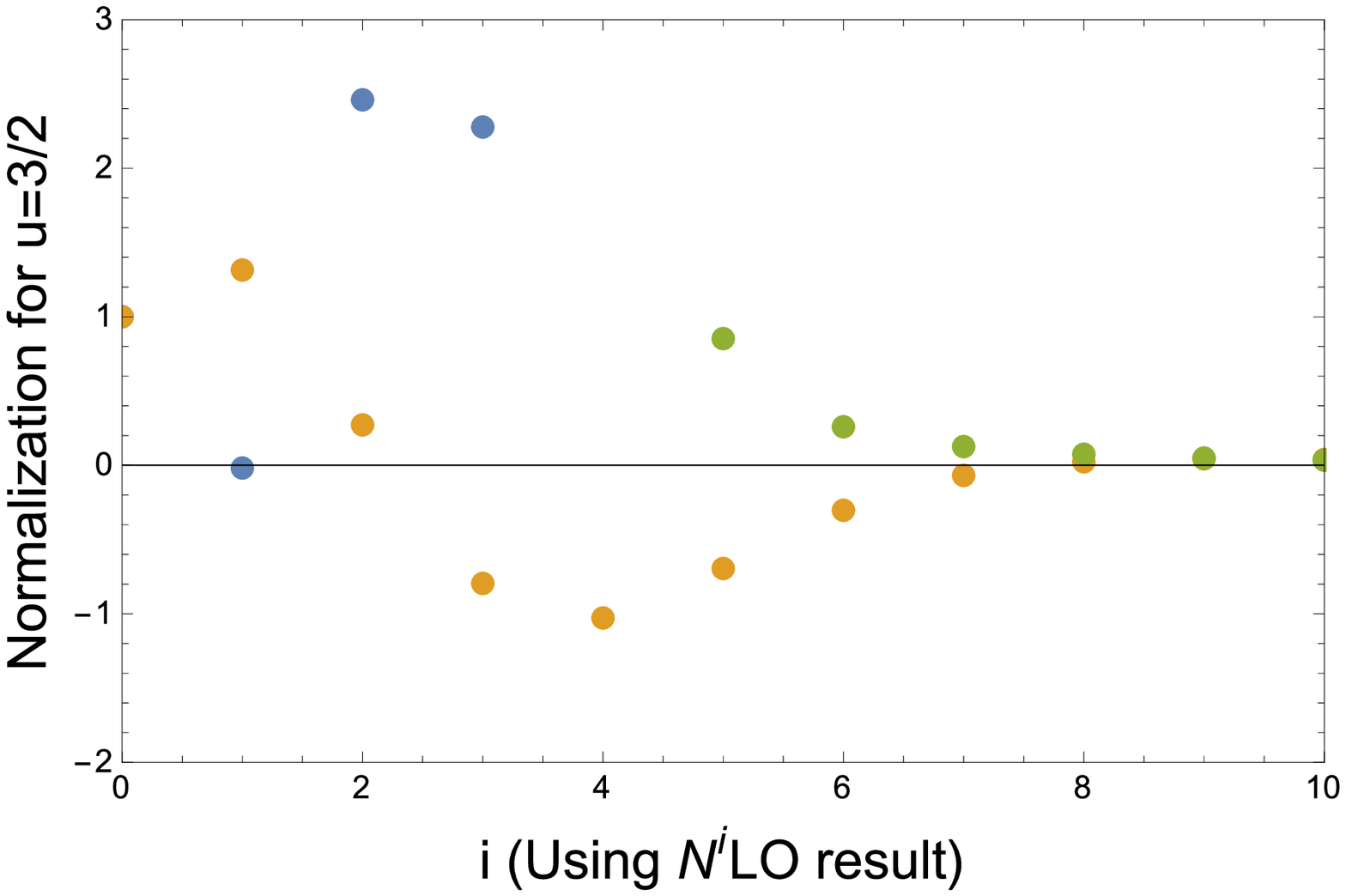}
\end{center}
\end{minipage}
\caption{\small
Estimates of the normalization constants of the renormalons at $u=1/2$ (left) and $3/2$ (right) for the momentum-space potential.
Green points are obtained from the RG-improved series,
while the orange points correspond to the large-$\beta_0$ approximation.
The first four points of the RG-improved series coincide with the
exact fixed-order results, shown by blue points.
In the right panel the $i=4$ point ($-7.30$) lies out of the plot range.
}
\label{fig:3}
\end{figure}

In Fig.~\ref{fig:3} we show the estimates of $N_{\widetilde{\alpha}_V, u=1/2}$
and $N_{\widetilde{\alpha}_V, u=3/2}$ from the RG-improved series, 
in addition to the ones from the fixed order results.
In these figures, we subtract the IR divergence in the
scheme (A) at NNNLO.
(Note that in the RG-improved series the terms beyond N$^3$LO
are zero for $\Vmos$ since we set $\mu=q$.)
We also plot their estimates using the large-$\beta_0$ approximation
for the higher-order terms (they are non-zero even beyond N$^3$LO).
In both cases we know that the normalization constants are zero.
We see in the figures that the estimates approach zero as
we include more terms.
Since the normalization constants are expected to be zero (even if we do not use any approximation), 
this figure shows overall consistency.

%
%

Thus, in both cases the observed results are consistent with
the expectation that the renormalon at $u=1/2$ is absent
and the $u=3/2$ renormalon is suppressed.
For $u=1/2$, we may already observe smallness of the renormalon contribution
from the known perturbative series.
For $u=3/2$, however, the number of terms are much too few to make
any statement on the size of the renormalon.
By using the formula \eqref{Vmom-renorm-absence} and
the fact that $r$-dependence of $V_A=1+{\cal O}(\alpha_s^2)$
is suppressed,
we can make a stronger prediction on the smallness of the renormalon.
We confirm validity of this formula using the higher-order
estimates by RG-improvement (trivial) or by 
the large-$\beta_0$ approximation.

\section{Conclusions}

We have investigated the $u=1/2$ and $u=3/2$
renormalons in the static QCD potential in position
space and momentum space.
In particular we have presented detailed examinations of the $u=3/2$ renormalon
for the first time.
In terms of pNRQCD EFT, 
we have studied the renormalon of the Wilson coefficient $V_S(r)$ 
(and in connection with this the second term of the multipole expansion, $\delta E_{\rm US}$, as well).

We have determined the structure of the $u=3/2$ renormalon based on the OPE (or multipole expansion) and the RG equations.
Although there are non-trivial features specific to the QCD potential 
(originating from the fact that the multi-scales are involved), we find that
the renormalon uncertainty can be parameterized (besides the overall normalization) 
similarly to the general case as reviewed in Sec.~\ref{sec:Theory}.
The relevant parameters are
the Wilson coefficient of the $\mathcal{O}(\vec{r})$ interaction $V_A$,
in particular its anomalous dimension
(associated with the logs from the soft scale), 
and also the coefficients of the beta function.
We have also clarified how the renormalon uncertainties of the position-space potential propagate
to the momentum-space potential.
The $u=1/2$ renormalon is completely absent in the momentum-space potential,
and the $u=3/2$ renormalon uncertainty is suppressed by $\alpha_s(q)^2$ 
in momentum space compared to that in position space.
While the renormalon uncertainty of the momentum-space potential has been believed to be small,
our result provides a quantitative insight on this issue.
We have 
given a systematic and precise analysis of the old problem,
including renormalization prescription and
treatment of the IR divergences (US logarithms) based on the multipole expansion 
in the pNRQCD EFT.

There are some difficulties caused by the IR divergences, however.
First, it is not obvious whether the renormalization of $V_S(r)$ to remove the IR divergences affects
the renormalon structure detected from the OPE argument. 
We have proposed a way to remove the IR divergences which is likely to keep the
renormalon structure unchanged based on our current knowledge.
Secondly, we have pointed out that it is difficult to eliminate the IR divergences
and IR renormalon at $u=3/2$ of $V_S(r)$ simultaneously in the multipole expansion, i.e., $V_S(r)+\delta E_{\rm US}$.
In particular, the perturbative result for the sum given by the double expansion in $\alpha_s$ and $\log (\alpha_s)$ 
is free from the IR divergences but not from the IR renormalon.
A systematic method which can subtract the IR renormalon as well
needs to be developed for obtaining an accurate prediction.
The contour deformation method 
used in Ref.~\cite{Takaura:2018lpw} has an advantage
in this respect (see below).

We performed numerical analyses
and checked our understanding as well as the current status of
our knowledge on the perturbative series of $V_S(r)$ and $\Vmos$.
With the available first four terms of the perturbative series,
we find that already the normalization constants of the $u=1/2$ renormalons
can be estimated with moderate accuracies (consistent with
the analyses \cite{Pineda:2001zq}).
On the other hand, the normalization constants of the $u=3/2$ renormalons
are still not reachable.
According to the analysis for the RG-improved series (neglecting beyond NNNLL terms),
it is suggested that we need 15-20 terms of the series expansion
to obtain reliable estimates of the normalization constant.
In the same RG method, we obtained an analytic formula for the normalization constants
for half-integer renormalons [eq.~\eqref{generalI}],
which is confirmed to be valid by comparison with the estimate using
Lee's method, which utilizes finite number of terms of perturbative series.

We noted the existence of a peculiar renormalon at $u=1/2$, 
which is related to IR divergences of the static QCD potential in naive perturbation theory
and induces an uncertainty of $\mathcal{O}(\LMS r^2 \Delta V^2(r))$.
This can be an obstacle in estimating the normalization constants of the familiar 
renormalons at $u=1/2$ and $u=3/2$.
To investigate the familiar $u=1/2$ renormalon (which induces an $r$-independent uncertainty),
it is better to study perturbative expansion of the pole mass in terms of the $\overline{\rm{MS}}$ mass,
which is free from IR divergence.
To study the normalization of the $u=3/2$ renormalon,
we proposed a method using a non-trivial mapping,
which is not disturbed by the renormalons at $u=1/2$.

As an application, the present work clarifies the status of
the method (contour deformation method) used in
a recent determination of $\alpha_s(M_Z)$ from $\Vpos$
after subtracting the $u=1/2$ and $3/2$ 
renormalons \cite{Takaura:2018lpw}.
(The IR divergence is canceled as well.)
There, it is assumed that the corresponding renormalons contained in
$\Vmos$ can be neglected.
As we have seen in Sec.~\ref{sec:3.4}, 
the normalization of
the $u=1/2$ renormalon in the momentum space potential is
exactly zero.
For the $u=3/2$ renormalon, it turned out that the dominant (or leading) uncertainty $\sim r^2 \LMS^3$,
which comes from the IR region of the Fourier transform of the momentum space potential, 
is subtracted since this method removes the IR region.
On the other hand, the subleading part $\sim r^2 \LMS^3 \alpha_s(1/r)$, which comes from
the uncertainty of the momemtum-space potential, is 
generally expected to remain.
This shows how
the $u=3/2$ renormalon
is suppressed theoretically, and
the current status is that
with the first four terms of the perturbative series
the $u=3/2$ renormalon in $\Vmos$ is far from detectable, 
based on the detailed numerical analysis.
Thus, we obtain the following overview:
(1) As demonstrated in Ref.~\cite{Takaura:2018lpw}, the contour deformation
prescription is indeed useful to raise accuracy of the prediction for
$\Vpos$ in the low energy region.\footnote{
This feature originates not only from subtracting renormalons
but also from removing an unphysical singularity from the prediction.
}
We have clarified how the assumption used in this prescription can be justified.
(2) At the same time, we still do not have a sufficient sensitivity to make
quantitative estimate of
the normalization of the $u=3/2$ renormalon in $V_S(r)$, and this is consistent
with the analysis in Ref.~\cite{Takaura:2018lpw}, where
the normalization of the $r^2$ term ($A_2$) in the OPE 
has an order 100\% error due to the uncertainty from 
unknown higher-order perturbative corrections.\footnote{
See footnote 21 in the second paper of \cite{Takaura:2018lpw}.
}
Therefore, the method 
is reasonable for steadily improving accuracy of 
$\Vpos$, by separating and
subtracting the renormalons from $V_S(r)$
using the currently known terms of the perturbative series.

In this paper we used the RG equations
of the soft scale $1/r$ (combined with OPE).
We believe that they determine the major structure of the
renormalons at $u=1/2$ and $u=3/2$ in the potentials.
It may also be useful to use the RG equations of the US scale
$\mu_f$ in order to study further the detailed structure of
the renormalons.
This was already indicated in the examination of the unfamiliar renormalon 
at $u=1/2$ in Sec.~3.5.
We leave it to future investigation.

To study renormalons beyond $u=3/2$, there still remain
works to be done.
In particular, it has not been clarified yet which renormalons are
specified from the OPE of pNRQCD EFT beyond $u=3/2$.

\section*{Acknowledgements}

The authors are grateful to Yuichiro Kiyo and Antonio Pineda for fruitful discussion.
The works of Y.S.\ and H.T., respectively, were
supported in part by Grant-in-Aid for
scientific research (Nos.\  17K05404 and 19K14711) from
MEXT, Japan.

\section*{\boldmath Appendix A: IR cancellation at $O(r^0)$}

\begin{figure}[t]
\begin{center}
\includegraphics[width=9cm]{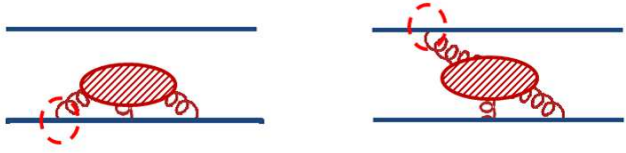}
\caption{\label{IR-cancellation}\small
To take into account the couplings of a gluon to
the total static currents, both self-energy
and potential-energy diagram contributions need
to be included, and
a cancellation takes place between them
in the IR limit of the gluon momentum $|\vec{q}|\to 0$.
}
\end{center}
\end{figure}

The cancellation of IR contributions between the
self-energy $2\Sigma$ and the potential energy $V_{S}(r)$
is a general property of gauge theory, which can be seen as follows.
A static current 
has only the time 
component,
\bea
j^\mu_{a,i}(x)=\pm T_a\delta^{\mu 0}\delta^3(\vec{x}\mp\vec{r}/2) ,
~~~~~(i=Q,\bar{Q})
\label{static-current}
\eea
since a static color charge has no spatial motion.
Here, $\pm\vec{r}/2$ denote the
positions of the static charges
$Q$ and $\bar{Q}$.
(We fix the c.m.\ coordinate to the origin $\vec{0}$.)
Hence,  an IR gluon, which couples to the static 
currents 
via minimal coupling $A_\mu^a(\vec{q},t)\, j^\mu_{a,i}(-\vec{q},t)=
A_0^a(\vec{q},t)\, j^0_{a,i}(-\vec{q})$,
 couples to
the total charge of the system in the IR limit $|\vec{q}|\to 0$:
\bea
Q_a^{\rm tot}=\sum_{i=Q,\bar{Q}} j^0_{a,i}(\vec{q}=\vec{0}) .
\eea
Therefore, an 
IR gluon decouples
from a static color-singlet system.
Diagrammatically, however, an IR gluon can
detect the total
charge of the system only when both self-energy 
diagrams
and 
potential-energy diagrams
are taken into account, as can be
seen from Fig.~\ref{IR-cancellation}.
This means that 
a cancellation takes place between these two types of diagrams,
since the IR gluon couples to individual diagrams
but decouples from the sum of them.

On the other hand, in analogy with classical electrodynamics,
gauge field couples to the total charge of the system
in the lowest order [$O(r^0)$] of the multipole expansion:
\bea
\int d^3\vec{x} \, A^a_\mu(\vec{x},t)\, J^\mu_a(\vec{x},t)
= A^a_0(\vec{0},t)\int d^3\vec{x} \,  J^0_a(\vec{x},t)
+ O(r^1)\,,
~~~~~
J^\mu_a=\sum_{i=Q,\bar{Q}}j^\mu_{a,i}
\,,
\eea
which follows from eq.~\eqref{static-current}.
Accordingly, in the pNRQCD Lagrangian (in the static limit),
there is no coupling of the singlet field $S$
and the gauge field at the lowest order of the
multipole expansion \cite{Brambilla:2004jw}.
Hence, the IR cancellation between the self-energy and
potential-energy diagrams is explicit at $O(r^0)$
in the multipole expansion (OPE) of the total energy of
a static $Q\bar{Q}$ pair.\footnote{
The  $O(r^0)$ part of $V_{S}(r)$, which is relevant to
the leading $O(\LQ)$ renormalon, is free of IR divergences.
It is consistent with the fact that
the pole mass is known to be IR finite at each order of the
perturbative expansion \cite{Kronfeld:1998di}.
As discussed in Secs.~3.3 and 3.5, IR divergences of $V_{S}(r)$
cancel against the $O(r^2)$ part and beyond in
the OPE of $V_{\rm QCD}(r)$.
The IR divergences [or more physically US logarithms of $V_{\rm QCD}(r)$]
are generated by color dipole and
higher multipoles of the static $Q\bar{Q}$ system.
}

Intuitively,
IR gluons with wavelengths of order $\Lambda_{\rm QCD}^{-1}(\gg r)$
cannot resolve the color charge of each particle, hence
they only see the total charge of the system.
More accurately,
coupling of IR gluons to the system can be expressed by
an expansion in $\vec{r}$ (multipole expansion)
for small $r$,
in which the zeroth multipole (=total charge) of
the color-singlet $Q\bar{Q}$ pair is zero.

The modern approach (after around 1998)
to use the $\overline{\rm MS}$ mass
for the computation of $E_{\rm tot}(r)=2m_{\rm pole}+V_{S}(r)$
for a heavy quarkonium system
can be viewed as follows.
The total energy of the system is computed as the sum of 
(i) the
$\overline{\rm MS}$ masses of $Q$ and $\bar{Q}$, (ii) 
contributions to the self-energies
of $Q$ and $\bar{Q}$ which are not included in the $\overline{\rm MS}$ masses,
and (iii) the potential energy between $Q$ and $\bar{Q}$.
Contributions of IR gluons with wavelengths larger than $r$
automatically cancel between (ii) and (iii) in this computation
\cite{Brambilla:2001fw}.
In this way we can eliminate a large part of
the IR contributions from the
computation of $E_{\rm tot}(r)$.

\section*{\boldmath Appendix B: Derivation of eq.~\eqref{resum2}}

We show some details of the derivation of eq.~\eqref{resum2}.
The regularized
dimensionless potential is given by 
\bea
v_+ &=& 
\int_0^{\exp(+ i \epsilon)\times\infty} dt\, e^{-t/\alpha_s}\,
B_{v}(t)
\nonumber\\ 
&=&
-\frac{2C_F}{\pi}
\int_0^{\exp(+ i \epsilon)\times\infty} dt\, e^{-t/\alpha_s}\,
\int_0^\infty \frac{dq}{q}\,\sin(qr)
B_{\widetilde{\alpha}_V(q)}(t)
\nonumber\\ 
&=&
-\frac{2C_F}{\pi}
i \int_0^{\infty} ds\, e^{-is/\alpha_s}\,
\int_0^\infty \frac{dq}{q}\,\sin(qr)
B_{\widetilde{\alpha}_V(q)}(is)
\,.
\eea
The integral contour of $t$ is rotated to the positive imaginary axis
($t=is$).
$v_-$ can be obtained by setting $\epsilon\to -\epsilon$
and $s\to -s$
(or, by taking the complex conjugate).

The Borel transform of $\widetilde{\alpha}_V(q)$ can be expressed
in the integral form as\footnote{
Expanding $\widetilde{\alpha}_V(q)$ in $\alpha_s=1/p$,
the integral at each order of the expansion can be evaluated easily
by the residue theorem.
Note that the integral contour of $p$ is closed in the upper-half plane
for $s>0$.
}
\bea
B_{\widetilde{\alpha}_V(q)}(is)
=\int_{-\infty-i\epsilon}^{+\infty-i\epsilon} \frac{dp}{2\pi i}\, e^{ips} \,
\widetilde{\alpha}_V(q)\biggr|_{\alpha_s\to 1/p}
\,.
\eea
We approximate $\widetilde{\alpha}_V(q)$ by
$ [\widetilde{\alpha}_V(q)]_{\text{N$^k$LL}} $.
According to our current knowledge of the RG equation  at N$^k$LL,
$\alpha_s(q)$ diverges at $q=q_*$ if the running starts from $\mu>q_*$
with the initial condition $\alpha_s(\mu)=1/p>0$.
In this case there is a singularity on the real axis of $p$,
due to the singularity of $\alpha_s(q)$.
At LL ($k=0$), the singularity is located at $p=b_0\log(\mu^2/q^2)$.
At N$^k$LL, the singularity of $p$ is on the real axis
for given values of $q(>q_*)$, $\mu$ and $k$, where
the relation is given by
$q=q_*(\mu,\alpha_s(\mu)=1/p;k)$.
We are concerned with the case that $q$ is in the vicinity of $q_*$, where
$\alpha_s(q)\sim (q-q_*)^{-1/(k+1)}$.
The singularity of $p$ 
can be shifted infinitesimally into the upper-half plane (hence, without
crossing the contour of $p$ integral) by a shift $q\to q-i\epsilon$  in the vicinity of $q_*$.

After changing the order of the
integration, we can integrate over $p$ and $s$, which transforms
$\widetilde{\alpha}_V(q)$ to $\widetilde{\alpha}_V(q)$.
Then we are left with the $q$ integration with the integral
contour deformed into the lower-half plane  in the vicinity of $q_*$.
The above $i\epsilon$-prescription for $q$ specifies how to avoid
the singularity of $ [\widetilde{\alpha}_V(q)]_{\text{N$^k$LL}} $ 
at $q=q_*$ compatibly with
the deformation of the integral contour of $t$.

\end{document}